\documentclass[acmsmall]{acmart}

\AtBeginDocument{%
  \providecommand\BibTeX{{%
    \normalfont B\kern-0.5em{\scshape i\kern-0.25em b}\kern-0.8em\TeX}}}

\setcopyright{acmcopyright}
\copyrightyear{2021}
\acmYear{2021}
\acmDOI{XXX}

\acmVolume{XX}
\acmNumber{X}
\acmArticle{XXX}
\acmMonth{4}
\usepackage{algorithmic}
\usepackage{graphicx}
\usepackage{textcomp}
\usepackage{multirow}
\usepackage{color}
\usepackage{gensymb}
\usepackage{caption}
\usepackage{subcaption}
\usepackage{url}
\usepackage{lscape}
\usepackage{dblfloatfix}
\usepackage{tabularx}
\usepackage{booktabs} 
\usepackage{longtable}
\usepackage{afterpage}
\usepackage{caption}
\usepackage{subcaption}
\usepackage[flushleft]{threeparttable}

\newcommand\newversion[1]{{\color{black} {#1}}}
\def\BibTeX{{\rm B\kern-.05em{\sc i\kern-.025em b}\kern-.08em
    T\kern-.1667em\lower.7ex\hbox{E}\kern-.125emX}}

\graphicspath{{Figures/}}
\DeclareGraphicsExtensions{.pdf,.png,.jpg,.eps} 

\begin{document}

\title{Hardware Trust and Assurance through Reverse Engineering:\\}
\subtitle{A Tutorial and Outlook from Image Analysis and Machine Learning Perspectives}


\author{Ulbert J. Botero}
\authornotemark[1]
\email{jbot2016@ufl.edu}
\affiliation{$^1$}
\author{Ronald Wilson}
\email{ronaldwilson@ufl.edu}
\authornote{Ulbert J. Botero and Ronald Wilson contributed equally to this research.}
\affiliation{$^1$}
\author{Hangwei Lu}
\email{qslvhw@ufl.edu}
\affiliation{$^1$}
\author{Mir Tanjidur Rahman} 
\email{mir.rahman@ufl.edu}
\affiliation{$^1$}
\author{Mukhil A. Mallaiyan}
\email{mukhil.mallaiyan@ufl.edu}
\affiliation{$^1$}
\author{Fatemeh Ganji}
\email{fganji@wpi.edu}
\authornote{Corresponding author. Fatemeh Ganji was with Florida Institute for Cybersecurity Research, University of Florida, when this study has been done.}
\affiliation{$^2$}
\author{Navid Asadizanjani}
\email{nasadi@ece.ufl.edu}
\affiliation{$^1$}
\author{Mark M. Tehranipoor}
\email{tehranipoor@ece.ufl.edu}
\affiliation{$^1$}
\author{Damon L. Woodard}
\email{dwoodard@ece.ufl.edu}
\affiliation{$^1$}
\author{Domenic Forte}
\renewcommand{\shortauthors}{Botero, Wilson, et al.}
\email{dforte@ece.ufl.edu}
\affiliation{$^1$}


\affiliation{%
  \institution{\\
  1 Florida Institute for Cybersecurity Research, University of Florida}
  \streetaddress{601 Gale Lemerand Dr}
 \city{Gainesville}
  \state{FL}
 \postcode{32611}}
 \affiliation{
  \institution{2 Worcester Polytechnic Institute}
  \streetaddress{100 Institute Road}
 \city{Worcester}
  \state{MA}
 \postcode{01609-2280}
}



\begin{abstract}
In the context of hardware trust and assurance, reverse engineering has been often considered as an illegal action. Generally speaking, reverse engineering aims to retrieve information from a product, i.e., integrated circuits (ICs) and printed circuit boards (PCBs) in hardware security-related scenarios, in the hope of understanding the functionality of the device and determining its constituent components. 
Hence, it can raise serious issues concerning Intellectual Property (IP) infringement, the (in)effectiveness of security-related measures, and even new opportunities for injecting hardware Trojans. 
Ironically, reverse engineering can enable IP owners to verify and validate the design. Nevertheless, this cannot be achieved without overcoming numerous obstacles that limit successful outcomes of the reverse engineering process. This paper surveys these challenges from two complementary perspectives: image processing and machine learning. These two fields of study form a firm basis for the enhancement of efficiency and accuracy of reverse engineering processes for both PCBs and ICs. In summary, therefore, this paper presents a roadmap indicating clearly the actions to be taken to fulfill hardware trust and assurance objectives.

\end{abstract}



\settopmatter{printfolios=true}
\maketitle

\section{Introduction}\label{sec:introduction}

Outsourcing of integrated circuit (IC) and printed circuit board (PCB) design, fabrication, packaging, and testing has dramatically reduced product development time and cost. In doing so, this has enabled the widespread availability of microelectronics, which has indeed transformed modern life. However, unintended consequences include malicious design alteration (i.e., hardware Trojan insertion \cite{robertson2018bloomberg,samuelk.moore2018}) and the rise of the counterfeit electronics industry~\cite{tehranipoor2015counterfeit}. Reverse engineering (RE) is widely applied for educational purposes and for detecting Intellectual Property (IP) infringement, but it can play an even more significant role in hardware trust and assurance. RE of electronic chips and systems refers to the process of retrieving an electronic design layout and/or netlist, stored information (memory contents, firmware, software, etc.), and functionality/specification through electrical testing and/or physical inspection. Although RE is often presented in a negative light (e.g., illegal cloning designs and/or disclosing sensitive information to a competitor or adversary), it is sometimes the only foolproof way to detect malicious alteration and/or tampering by semiconductor foundries, find vulnerabilities present in commercial-off-the-shelf (COTS) chips and avoid them, and replace obsolete (i.e., no longer manufactured) hardware.

As for attaining trust and assurance, existing techniques are limited and/or ineffective. For example, run-time monitoring techniques increase the resource requirements -- power consumption, memory utilization, and area overhead on ICs/PCBs -- due to on-chip/board sensors used to detect anomalous activities. In test time methods, the challenge is to generate test vectors that trigger stealthy, well-placed hardware Trojans in billion-transistor chips. Similarly, in side-channel signal analysis approaches, inescapable process variations, and the measurement noise undermine the probability of detecting small Trojans~\cite{jin2008hardware}. As a result, the confidence level in detecting Trojans using the aforementioned techniques is quite low~\cite{narasimhan2011tesr,chakraborty2009mero, tehranipoor2010survey}.  Hence, RE has been gaining more attention in recent years and experiencing community-wide acceptance as an effective approach, in particular, for hardware Trojan detection~\cite{courbon2015high,bao2016reverse}. 

In the area of IC counterfeit detection and avoidance, the current best practice requires the use of either classification by subject matter experts (SME), procuring lifetime buys for long-term system maintenance, or acquiring components from untrusted distributors in a supply chain, which potentially involves grey market distributors. Each of these options is non-ideal. The large quantities of components that SME counterfeit analysts are required to analyze and manually classify makes this current practice very inefficient and costly. As for life-of-type buys, it is impractical and almost impossible to predict the lifetime of every component in a design, in anticipation of obsolescence and failure. Overestimation of the lifetime leads to procuring more components than necessary, and consequently, the waste of resources. Underestimation of the lifetime results in non-ideal situations, such as redesign or procurement through grey market distributors necessitated earlier than desired.

For PCBs, counterfeiting and Trojan insertion is a similarly prevalent problem. While there are existing chip-level integrity validation approaches, as mentioned above, they are not readily adaptable to PCBs, which is a cause for concern. In response to this concern, a common method for preventing and protecting against PCB counterfeiting is to take advantage of intrinsic characteristics of PCBs making each and every of them (quite) unique \cite{zhang2015robust}. Additionally, \cite{iqbal2017pcb} has explored using unique patterns seen in images of surface vertical interconnect access (via) as fingerprints of design to overcome the problem of counterfeit PCB distribution. While both of these approaches can help us to improve reliability and assurance of a PCB after manufacturing, these techniques would still have to face difficulties in detecting small Trojans, similar to that seen in the October 2018 Bloomberg Businessweek article, entitled "The Big Hack" \cite{robertson2018bloomberg}. In October 2018, it was claimed that unauthorized microchips were found in the products of a manufacturer that provided Apple, Amazon, and even the US government with specialized servers \cite{samuelk.moore2018}. As reported in \cite{samuelk.moore2018}, security experts suspected that the assembly facility owned by Supermicro might have implanted the chip, which could serve as a backdoor for spying information exchanged over networks equipped with the altered PCBs of servers. Such an attack, i.e., adding an extra chip maliciously, severely affects the confidentiality and integrity of a system. More importantly, the survivability of this system is strongly influenced due to the typically high degree of complication and obstacles involved in revealing the existence of such threats and recovering the system from them. This can further highlight the strong demand for the verification of the security of the physical systems.

According to the above discussion, today more than ever, there is a significant need for fast and fully automated RE, imposed by industries, and especially for security-critical applications. The RE process comprises delayering, imaging, annotation, and netlist extraction. The current state-of-the-art practices are tedious, challenging, and expensive. They usually require a suite of cleanroom and microscopy equipment, very long imaging times, and manual or semi-automated post-processing steps for converting images to netlists. 
Despite this, recent advancements in failure analysis tools and delayering processes are opening up new dimensions in RE. As an example, plasma etching has achieved better control over ion-energy distribution, thereby improving selective and automation in delayering \cite{rahman2014analytical}. Furthermore, the introduction of non-destructive X-ray computed tomography (X-Ray CT) and ptychography in recent years can eliminate the process of delayering, and hence, can speed up the imaging time for the upper metal layers of an IC and an entire PCB. New scanning electronic microscopes (SEMs), such as multi-beam systems, have also been introduced to significantly speed up imaging of nanoscale samples. Nevertheless, they are not widely available and are still several times more expensive than standard SEMs. In addition, since such tools could yield petabytes of data in only a day, the research on automated and intelligent image analysis algorithms is an urgent need to reduce the time and cost of RE. 

In this tutorial, we systematically study the current challenges that automated RE faces in order to be useful for providing trust and assurance. Existing surveys on RE focus on different aspects, e.g., Keshavarz et al. have presented examples of image-based RE applications and discussed hardware attacks in detail \cite{keshavarz2018survey}, while Fyrbiak et al. have summarized the process of accessing gate-level netlist from three system models and discussed the evaluation strategies \cite{fyrbiak2017hardware}. Compared to our work, they have explored neither the challenges during a typical RE process from an imaging perspective nor considered the possibility of applying machine learning approaches in this context. Our tutorial further describes a typical workflow of RE, and then investigates the possibilities and limitations of processes incorporated in such a workflow from the RE perspective. More precisely, we explain inherent differences between natural images, which virtually all the well-developed image processing algorithms are designed for, and images taken to conduct RE on a hardware device. To this end, we give an exhaustive overview of numerous obstacles to applying common methods originating in image processing and machine learning. In particular, we place emphasis on the need to incorporate domain knowledge to overcome them. Several examples of such issues are given and reviewed in detail. In summary, this tutorial aims at providing an outlook on how to improve RE so that it can better handle detection and avoidance tasks in the context of hardware trust and assurance. 

\begin{figure*}
    \centering
    \includegraphics[width=1.0\linewidth]{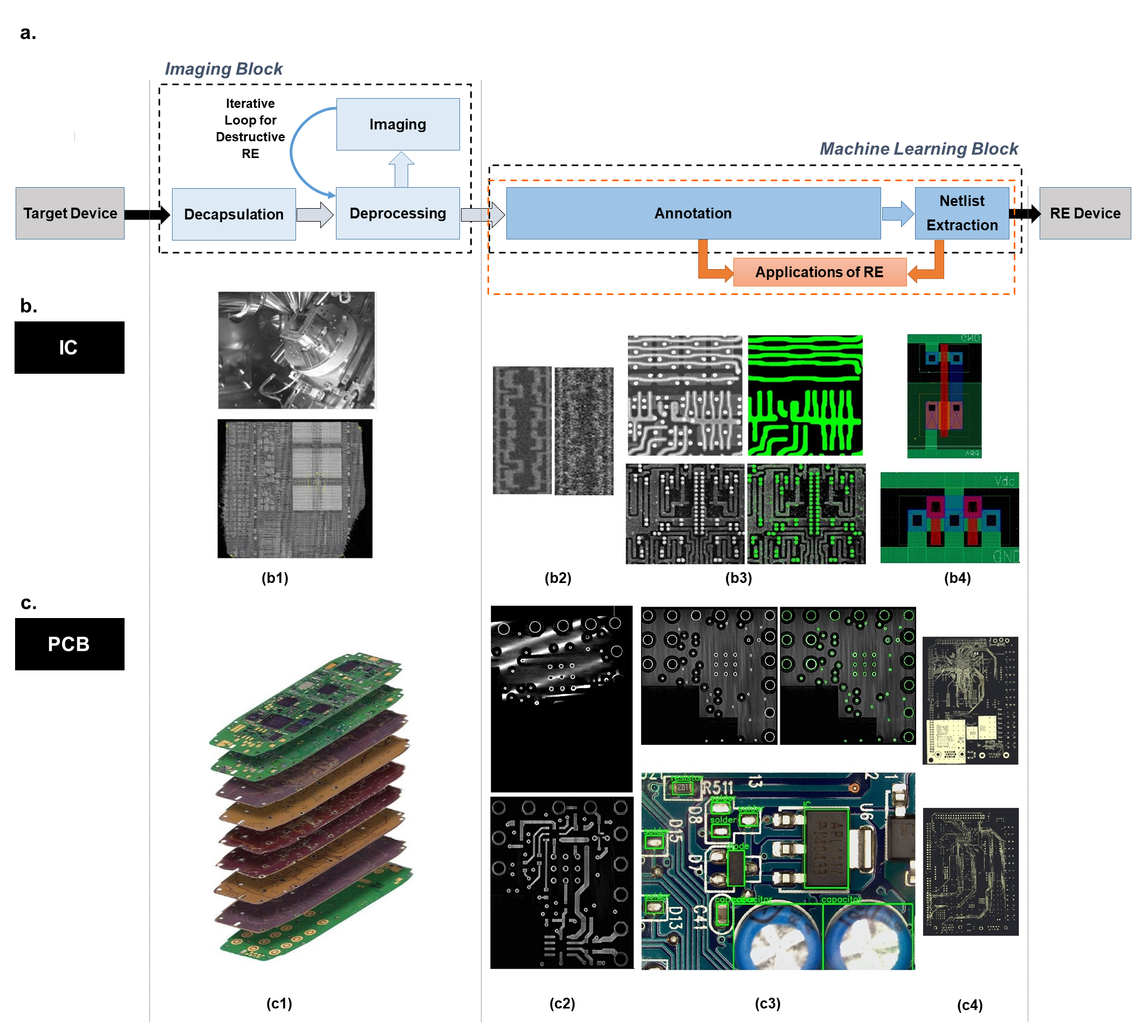}
    
    \caption{Our systematic overview of an RE process, which can be performed on ICs and PCBs, its challenges and possibilities. (a) A typical workflow of RE encompassing various stages. Two main blocks of such a workflow are: Image Analysis (see Section \ref{sec:section2}) and Machine Learning (see Section \ref{sec:section3}). Moreover, we discuss how the outputs of the machine learning-related block can enable us to provide hardware-based trust and assurance, as an application of RE (for a general view, see Section \ref{sec:section2-trustassurance}). Inherent challenges facing us in both cases of ICs and PCBs are further discussed in Section \ref{sec:section2}- \ref{sec:section3}. 
    (b) RE workflow for IC: (b1) Deprocessing of the IC \cite{principe2017steps}, (b2) Example of noise removal in the active region using different imaging parameters, (b3) Segmentation and extraction of polysilicon structures and vias in an IC \cite{cheng2018hierarchical,cheng2018hybrid}, (b4) Netlist of extracted logic cells. (c) RE workflow for PCB: (c1) Image depicting a multi-layered PCB \cite{torrance2009state}. Depending on the number of the layers in a PCB, different types of RE techniques should be considered. Irrespective of this, these challenges are inevitable: (c2) Example for misaligned layer and reconstructed image, (c3) Segmentation and extraction of vias for X-rayed PCB and labelled components on the surface of an optically imaged PCB, (c4) Segmented  layout  of  PCB  layers with  connected and not-connected vias \cite{asadizanjani2017pcb}. }
    \label{workflow}
\end{figure*}

\vspace{10pt}
\noindent\textbf{A brief overview and the organization of the tutorial: }Beyond providing a taxonomy of approaches proposed to address trust and assurance issues, Section 2 describes how automated RE can enable us to solve those problems more effectively. Section 3 provides a high-level overview of the challenges with RE. Section 4 discusses the issues impeding the RE workflow from an imaging perspective (see the imaging block in Figure \ref{workflow}). 
This section is complemented by Section 5 with an in-depth discussion from a machine learning and image analysis point of view along with a brief discussion on the application of deep learning in RE, as illustrated in the machine learning block in Figure \ref{workflow}. Section 5 further demonstrates how various applications of RE, such as counterfeit and Trojan detection, can leverage the information retrieved through feature extraction and analysis. 
Section 5 also briefly discusses counter RE approaches implemented in contemporary hardware. 
As this tutorial aims at pointing to a new outlook, Section 6 is devoted to future research directions. Finally, we conclude by pointing out the issues addressed in the tutorial. 
\section{Applications of RE for Trust and Assurance}\label{sec:section2-trustassurance}
Semiconductor technology has become an integral part of our everyday lives, as ICs and embedded systems have been becoming ubiquitous.  
The spectrum of the applications of these devices and systems covers various areas including, but not limited to, household appliances, critical infrastructures (i.e., commercial facilities sector, government facilities, energy sector, etc.), and military systems.   
Regardless of these applications, their trustworthiness and reliability must be assured. 
This section aims to explain how automated RE can address this concern by providing an added degree of precision for the analysis and evaluation, applied at different development stages in electronics industries. We further elaborate on the applications of (automated) RE, namely Trojan detection and obsolescence replacement.
\subsection{Trojan Detection and Counterfeit Avoidance}\label{sec:trojanCounterfeitApp}

Counterfeit and tampered electronics pose serious threats to hardware-based trust and assurance. In particular, cloned chips and hardware Trojans can violate the security requirements of root-of-trust, thereby reducing confidentiality, integrity, and availability. For ICs, cloning is the process of copying and unauthorized production of a design without having legal IP rights. Moreover, any malicious modification of the structure, functionality, or parameters of the chip that causes the device to operate outside of its specification can be identified as a hardware Trojan. 
Furthermore, the root-of-trust can be compromised 
at the system level. PCBs give another opportunity for an attacker to tamper, clone, counterfeit, and insert a hardware Trojan. In fact, since PCBs lie at the heart of an electronic system and integrate several components to achieve the desired functionality, it is increasingly important to guarantee a high level of trust and reliability at such an integration stage. The aforementioned incident allegedly at Supermicro serves as an example (see Section \ref{sec:introduction}). 
Advances in the RE automation process can enable us to shorten the time to identify these types of threats at multiple levels of an electronic system \cite{asadizanjani2017pcb, rahman2018physical}. 

\begin{figure}[t]
\centering
\includegraphics[width=.5\linewidth]{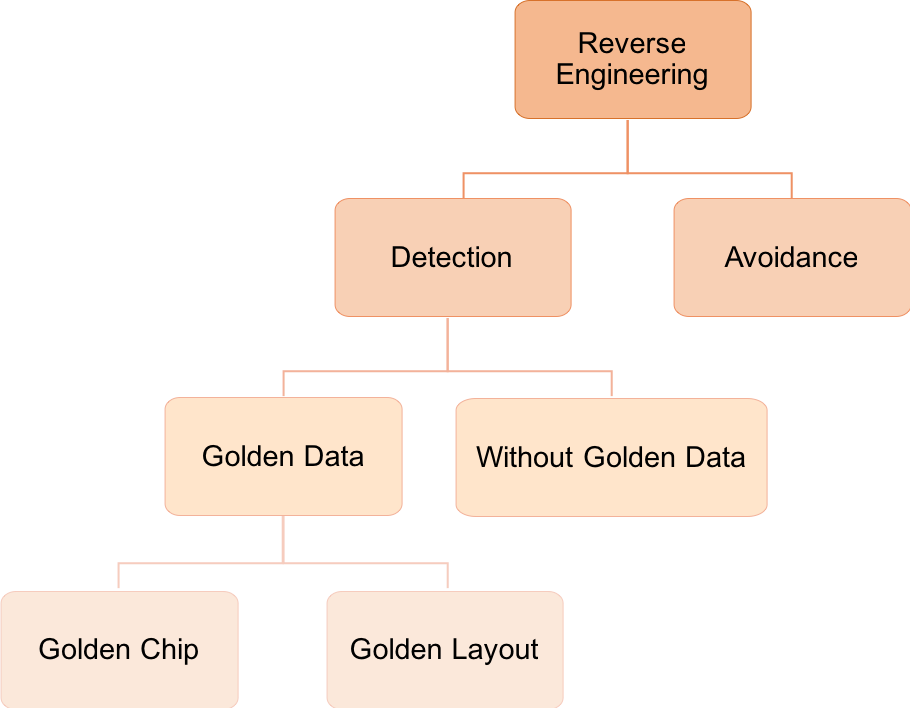} 
\caption{Taxonomy of approaches for addressing trust and assurance issues through RE.}
\label{fig:Taxonomy_trust}
\end{figure}

The importance of applying RE for addressing trust and assurance-related issues are twofold, namely detection and avoidance (see Figure \ref{fig:Taxonomy_trust}). 
When it comes to avoidance, we are interested in approaches that can prevent counterfeit parts from entering the supply chain. For this purpose, it is crucial to develop relatively less costly and time-consuming counterfeit detection methods \cite{guin2014counterfeit}. 
Therefore, due to this close connection between avoidance and detection, in this tutorial, our primary focus of interest is detection methods. In the detection process, the incoming electronic components undergo a physical or electrical inspection process to examine authenticity. As RE is an interior, physical-inspection-based approach, to decide whether a chip/system is cloned or to detect a Trojan, one should rely on the availability of golden data. Golden data can be images from a known authentic chip or PCB, bill of materials (BoM), schematic, layout, or device, whose functionality, structural and electrical parametric signatures are available for comparison. 

\begin{figure}
    \centering
    \includegraphics[width=.6\columnwidth]{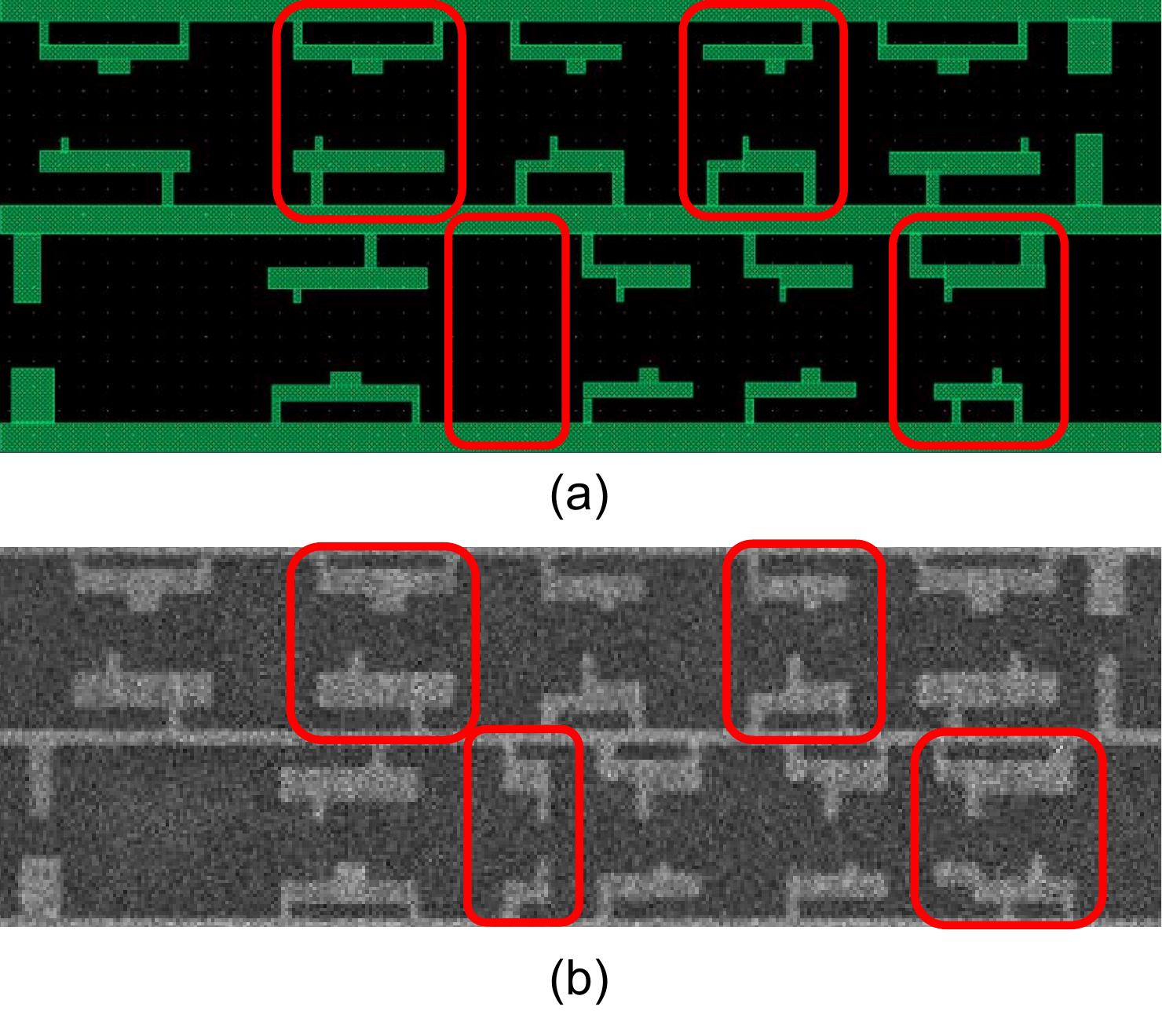}
    \caption{(a) Layout of smart card chip (b) SEM image of
the corresponding area. (Circled areas shows the effect of a modification or
insertion of the logic cell.)\cite{vashistha2018detecting}}
    \label{fig:icTrojan}
\end{figure}

For example, a layout\footnote{After performing the translation of a specification into a behavioral description (typically in a hardware design language (HDL)), this description is synthesized to generate a design implementation of logic gates, i.e., netlist. This netlist is used to produce a layout (GDSII file) by conducting placement/routing. To fabricate ICs,  this GDSII is sent to a foundry by the design house. } is determined as \textit{golden} if the IP holder and System-on-Chip (SoC) designer/PCB manufacturer are authorized and trusted\footnote{A trusted party is defined as one committed to ensuring a proper IC design/fabrication flow (i.e., does not insert Trojans, protects IP confidentiality, etc.). An untrusted party cannot ensure such a proper flow or performs malicious activities intentionally.} \cite{vashistha2018detecting}. A golden layout or design can provide a benchmark for assessing the functionality of the chip or analyzing its physical structure. The designer's layout (see Figure \ref{fig:icTrojan}(a)) can be compared to the SEM image taken from the respective manufactured design (see Figure \ref{fig:icTrojan}(b)), to determine possible Trojan insertions. However, it does not provide any reference for side-channel parametric profiles, e.g., power, path timing, electromagnetic signature, photonic emission, etc., which can only be characterized by using a fabricated chip or board. 
Additionally, a device is considered \textit{golden}, when either it is fabricated from a golden layout in a trusted facility or its functionally and physical characteristics are verified through full-blown RE \cite{tehranipoor2010survey}. The primary concern regarding fabricating a golden sample in a trusted facility is a prohibitively costly process. Besides, the parametric profile of the golden device is different from the same parametric profile of devices produced in another facility for the same technology node, even within the same fabrication facility. 

 Nevertheless, common test methodologies may not always be helpful for detecting Trojans \cite{jin2009experiences}. In this context, an RE approach can also be employed to detect extra insertions and deletions \cite{vashistha2018trojan, shi2019golden}.  
Note that although IC camouflaging\footnote{A technique that can be employed to mask the circuit functionality by synthesizing circuits with logic cells, which look similar, but can have different functionalities. }, especially dummy contact-based IC camouflaging, can impair the effectiveness of \emph{malicious} RE of ICs, the designer can greatly benefit from an automated RE along with a golden chip or layout to deal with such cases. 
Yet the challenges with RE-based approaches are the SME involvement and the execution time (see Section \ref{sec:obsolescenceApp}). 

For PCBs, due to the minute details involved in the Trojan insertion process, the availability of golden data to facilitate full-blown PCB RE is even more pressing. The modern nature of PCB designs, being multi-layered, provides a variety of Trojan insertion possibilities that are nearly impossible to prevent without full-blown RE. Specifically, an attacker can take advantage of unused pins, multiple layers, and hidden vias in the design to alter connections throughout the internal layers/ hidden vias, as well as the properties of these connections. Altering traces in the internal layers can make no structural difference, but produces undesired functionality under certain conditions. Such alterations include modifying the mutual coupling capacitance, characteristic impedance, and loop inductance \cite{ghosh2014secure} as well as adding ultra-low areas, and power components in the internal layers. Moreover, the chances of detecting these modifications via exhaustive testing are low since malicious functions are barely triggered during in-circuit and boundary-scan-based functional tests. 
With a full-blown RE, the design dimensions going down to the trace widths and spacing can be extracted and compared for tamper detection and to give the IP holder an available golden sample.  
If the attacker alters the design structure, by comparing the designed and extracted netlists, the detection can be less challenging, and the full-blown RE can be more effective (for more details see Section \ref{sec:obsolescenceApp}).        

In general, the detection techniques have progressed at a fast pace, due in part to advancements in artificial intelligence, and in particular, machine learning. 
Techniques originating from machine learning have been widely employed in hardware security. 
For instance, machine learning algorithms have been applied for Trojan and IC counterfeit detection; for a comprehensive survey, see~\cite{elnaggar2018machine}.  
Nevertheless, when it comes to approaches leveraging the strengths and capabilities of both reverse engineering and machine learning methods, e.g.,~\cite{bao2014application, bao2015reverse}, less effort has been made to develop such approaches.  
Only recently, as a result of the advancements in image analysis incorporated with the developments of techniques relying on SEM, X-Ray CT, and optical imaging, more reliable, faster and automated hardware Trojans detection methods have been developed, being also useful for detecting cloned chips/systems. 
Such a process generally involves several steps, namely image pre-processing, feature extraction, and classification. 

Image pre-processing influences the accuracy of perceptual feature extraction through noise reduction, edge enhancement, segmentation, etc. As the name implies, feature extraction deals with extracting salient features from the images of in the electronic component, acquired by using the SEM/X-ray CT/optical microscope. Those features are represented as inputs for machine learning algorithms, e.g., neural network, support vector machine (SVM) or clustering approaches, which can determine modifications in the function or the structure in the system. However, to benefit from advances in machine learning, relatively large sets of data are necessary to train machine learning algorithms. Especially for deep learning methods, a vast number of data samples are required to achieve an acceptable level of performance.
Nonetheless, advanced methods, e.g., Trojan Scanner \cite{vashistha2018detecting}, can direct trust and assurance-related studies towards partial RE-based hardware Trojan detection methods. 
 
In the presence of data derived from a golden sample, different methodologies, e.g., the structural test comparison between a suspected sample and the golden sample/layout, can be deployed to identify cloned devices \cite{guin2014counterfeit}.  Over the years, to address the availability of neither a golden chip/layout nor a sufficiently large training dataset when dealing with protecting chips/systems, different avoidance methodologies like the secure split-test, physically unclonable functions, and lightweight on-chip sensors have been proposed for countefeit detection and avoidance\cite{guin2014counterfeit}. In line with this, the fast and automated RE can enable us to establish a secured supply chain comprised of a trusted manufacturing facility and distribution for the security-critical applications. Such improvement offers effective measures for the avoidance of cloned or Trojan-infected chips.
 

\subsection{Obsolescence}\label{sec:obsolescenceApp}
In addition to Trojan detection and counterfeit avoidance, an RE-based method also provides trust and assurance for the obsolete or near-term life technologies and components. These technologies, usually referred to as legacy electronics/systems, are prominent in many critical systems. Typically the production cycle for electronics is under pressure from the fast-paced consumer electronics industry, where the next generation of devices with improved properties is expected and adopted  in the course of the following calendar year. Yet, the opposite is the case in military and government electronic systems that go through longer development cycles and deployment. These systems are designed to be in operation for decades \cite{stogdill1999dealing}. 
However, since these systems are deployed for increasingly longer periods, the cost of maintenance begins to increase due to needed parts becoming obsolete. The long life span of these components and systems opens up new possibilities for malicious activities including security concerns and vulnerabilities. Most notably, diminishing manufacturing sources for obsolete components can force original equipment manufacturers (OEMs) to purchase from untrustworthy distributors. This has been identified as a known source for recycled, remarked, or counterfeit components/systems and consequently, a pressing concern for governments, as reported by, e.g., the United States Senate \cite{senate2012inquiry}. 

Although a full system redesign is an option to address this concern, it is impractical due to the associated costs and manpower \cite{stogdill1999dealing}.
In particular, if previous design information that would be used for the redesign is no longer available or scarce, performing RE to acquire the needed design information can result in destroying the only available samples. 
This is often the case in legacy systems, where previous designs are lost over time through company migrations/transitions or components are obsolete and discontinued. These concerns are present for both of the IC and PCB levels, but can be addressed thanks to advances in image analysis and machine learning. 

\subsubsection{IC Level Upgrades}\label{sec:upgradeIC}
As an interdisciplinary field including several key components from image analysis and machine learning fields of study, automated RE enables us to replace obsolete technologies and provide additional trust and assurance in hardware security. With respect to the ability of automated RE to segment, identify, and interpret different properties of IC layouts, it is possible to not only deconstruct the netlist of a device, but also reconstruct it. By identifying various components on a layout and comparing them with the standard cells, the functionality and netlist of an IC can be deconstructed. Afterward, this information can be used either to analyze possible faults in the layout or for reproduction, if the reverse-engineered device is obsolete and no longer in distribution. Furthermore, once the functionality is deduced and the netlist is reconstructed, any desired upgrades (additional logic, security primitives, etc.) can be added to the design and the new upgraded design and layout are ready for fabrication \cite{botero2019upgrade}. 

\subsubsection{PCB Level Upgrades}\label{sec:upgradePCB}
The above-mentioned advances enable us to offer the maintenance or replace obsolete or rare PCBs as well. In a similar fashion to ICs, automated RE can be used on PCBs to identify key components, traces, vias, and layers to reconstruct the design and netlist \cite{mcloughlin2008secure}. Coupling these techniques with advances in non-destructive RE via X-Ray CT \cite{asadizanjani2017pcb} leads to an all-encompassing process that completely removes the traditionally needed SME. This is especially useful for PCBs, whose design information has been mishandled or with scarce supply. This can be explained by the fact that traditional RE may result in the destruction of samples undergoing the process \cite{grand2014printed}. Providing a substantial cost and efficiency savings achieved through this gathered design information, it is now possible to perform design-to-manufactured product validation, product-to-product comparison, and the ability to upgrade past designs. All of these provide an added level of trust and assurance to the systems that require the utmost attention to security.    



\subsection{Limitations of RE}

Upon completion of the workflow shown in Fig.~\ref{workflow}, gate-level connectivity, i.e., netlist, of the IC, can be extracted. However, the design hierarchy and the intend of the circuit blocks are still missing from the netlist. RE from the gate-level netlist to abstraction level - register-transfer level (RTL) and behavior (architectural) level is not a straightforward task since at the pre-silicon stage, each design undergoes extensive optimization performed by computer-aided-design (CAD) tools. CAD tools perform its optimization at the gate-level netlist by flattening the netlist, which results in loss of high-level design data, i.e., hierarchy and module information. Further, logic synthesis for logic minimization, technology mapping, sharing logic, and functional blocks increase the complexity in extracting RTL and behavior abstraction. In recent years, several formal verification methods, for example, Boolean Satisfiability Function (SAT)~\cite{subramanyan2013reverse}, Binary decision diagram (BDD)~\cite{xu2017novel}, and Quantified Boolean Function (QBF)~\cite{li2013wordrev}, have been used to identify the control logic and generate finite-state machine transition graph for extracting functional modules, e.g., registers, multiplexers, or modules like Arithmetic-Logic Unit (ALU), interfacing circuit, from the extracted GDSII file of the device. 

Alike behavior extraction of an IC, information retrieval from non-volatile memory (NVM) and physical unclonable functions (PUFs) is challenging. The NVM, such as Flash and EEPROM, stores charges as a medium of keeping the information safe during the power-off state of the device. However, reverse engineering the FLASH and EEPROM without disturbing the charge, even with the most sophisticated FA tools, is proved to be difficult~\cite{courbon2016reverse}. Unlike charge-based NVM, one-time programmables like E-fuse and anti-fuse can be reverse engineered using SEM and transmission electron microscopy (TEM)~\cite{rahman2020defense}. Though PUF is considered as a secured key-storage, the key stored in PUF is considered lost during destructive RE.    
\begin{figure*}[t]
    \centering
    \includegraphics[width=\textwidth]{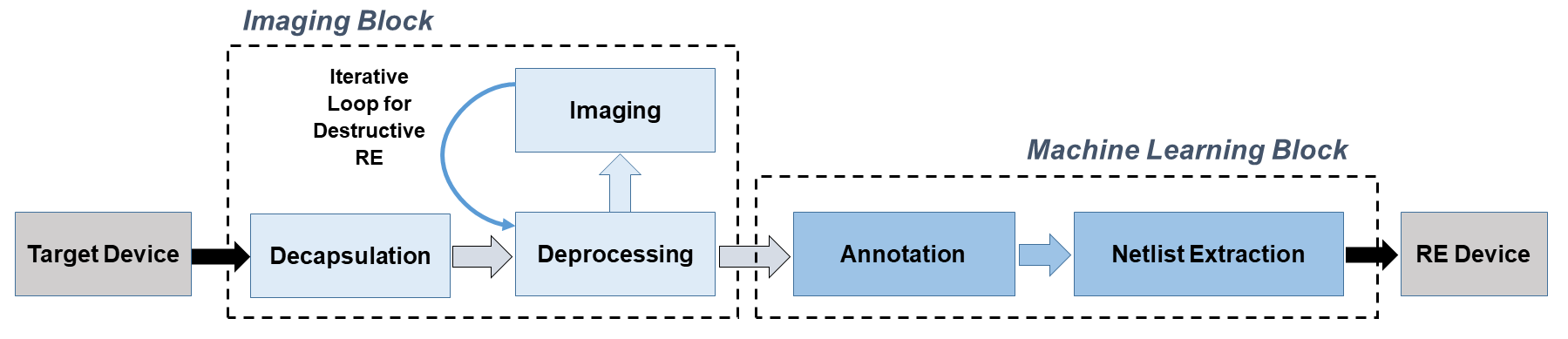}
    \caption{A typical RE workflow with demarcations for both image analysis and machine learning applicable blocks}
    \label{fig:re_flow}
\end{figure*}

\section{Reverse Engineering and associated challenges}\label{sec:sup}
Research and development in most domains are typically fueled by academic interest, commercial gain, and governmental entities for national interest matters.  With RE being opposite to the mainstream workflow for hardware design and manufacturing, research interest in it has been minimal. An exemplary instance of this fact could be observed in technical surveys with less focus on complete RE, in contrast to one of most relevant domain, i.e., applications of ML in hardware security \cite{elnaggar2018machine}. In addition, for hardware assurance, several studies considered RE as a last resort due to its overall complexity and costs. With the recent commercial interest in RE, stemming from the need for IP protection, it is necessary to devote more effort to research. 
In this regard, Lippmann et al. have suggested that given enough time and resources, RE can recover the building blocks of any given IC with up to 99\% accuracy \cite{lippmann2019integrated}. This estimate includes accounting for the possibility of countermeasures designed to thwart the performance of RE. From the perspective of commercial entities associated with hardware development and design, the goal is not to get designs to be completely RE-proof but to delay the process long enough to introduce completely novel designs and technologies. Due to this inherent conflict of interest, the amount of data available is limited as compared to other fields and, hence, limits the effectiveness of data-driven approaches such as machine learning in RE. 
Furthermore, compared to other relevant domains, e.g., failure analysis, techniques applied in the context of RE could not keep pace with the evolution of technology. 

To address these, RE should be performed as a highly modular process with a well-established workflow with a typical, practical example shown in Figure \ref{fig:re_flow}. This workflow has been tried and tested in a variety of works \cite{courbon2015high, torrance2009state, quadir2016survey, jain2017hardware, sarkar2019automating}. Typically, the development of a complex systematic approach, like RE, is initiated by drawing up a high-level overview of the problem. The lower level details are separated into modules for independent research and development depending on their overall contribution to an effective solution for the problem. However, in the case of RE, this process is reversed. 
For instance, imaging for ICs was predominantly used for fault and yield analysis in the forward manufacturing workflow, but was later adapted for RE. As a result of such an adoption,  development of modules have become unbalanced, which leads to process bottlenecks and missing links between established modules. The sequential nature of the workflow further exacerbates this problem, i.e., the errors accumulate at each layer, making the final result erroneous and unsuitable for the intended purpose.

To make the RE workflow more clear to the readers, the next sections elaborate on existing approaches associated with each module in the RE workflow, their inherent assumptions, potential drawbacks, and, finally, avenues of further research and development for the hardware assurance community. 

\section{Inherent Challenges Associated with Imaging Electronic Components and Systems}\label{sec:section2}
As depicted in Figure~\ref{fig:re_flow}, the imaging block encompasses all the steps involving physical interaction with the ICs and PCBs providing an effective demarcation between the hardware and software aspects of RE. The imaging block comprises of three key steps of the RE process: decapsulation, deprocessing, and imaging. 
This section describes these steps in detail for ICs and PCBs. 
\subsection{Problems Associated with Handling IC Images}\label{sec:image_IC}
The first step in the RE workflow is called decapsulation, taken to remove the protective enclosure surrounding the silicon die. The process of removal typically includes mechanical force and etching with acid. The various tried and tested recipes, like fuming nitric acid, can be found in literature \cite{skorobogatov2011physical}. 
This is then followed by deprocessing, and imaging steps, as explained below. 

\subsubsection{Deprocessing}\label{subsubsec:deprocess}
This is a decisive step in the RE workflow. Depending on the imaging modality available, the sample preparation varies. There are two major approaches: non-destructive and destructive imaging. 

\begin{figure}
    \centering
    \includegraphics[width=0.85\columnwidth]{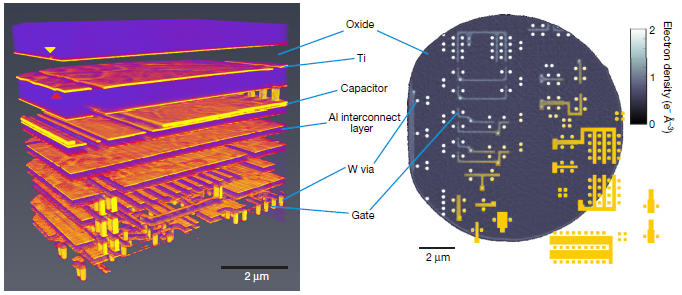}
    \caption{Non-destructive imaging using X-Ray CT \cite{holler2017high}}
    \label{fig:ct_recon_nature}
\end{figure}

\vspace{5pt}\noindent\textbf{Non-destructive imaging}: In the context of electronics, this type of imaging is predominantly used for fault analysis. In this regard, scanning confocal electron microscopy has been used to image buried structures in thick IC samples \cite{frigo2002submicron}. X-ray based imaging has been used for counterfeit IC detection and fault analysis \cite{mahmood2015real, lau2006non, levine1999tomographic, wang2011situ}. Scanning optical microscope has been used for counterfeit IC detection, but this approach assumes access to IC design files \cite{matlin2014non}. However, a recent study shows promising results for RE \cite{holler2017high}. The X-ray computed tomography (CT) technique has been able to effectively resolve features in an IC up to 14.6 nm. The CT reconstruction and a slice from the CT is shown in Figure \ref{fig:ct_recon_nature}. 

However, this advantage comes with significant overhead in the image acquisition time frame. With advancements in X-ray imaging, non-destructive may become a significant tool for RE. All the methods stated in this section are non-invasive. On a similar note, imaging techniques such as photon emission microscopy have been used for probing and decoding functionality of live ICs in near-infrared (NIR) spectra. Although these methods do assist in RE and have been widely used for hardware assurance purposes, they cannot be exploited for full-scale RE.  

\vspace{5pt}\noindent\textbf{Destructive imaging}: A wide range of electron microscopy techniques is employed in destructive imaging. Unlike imaging techniques presented in the previous section, these techniques require extensive sample preparation. Images acquired using improperly processed samples can have a detrimental impact on the final result. Electron microscopy techniques such as Scanning Electron Microscopy (SEM), Transmission Electron Microscopy (TEM), Focused Ion Beam (FIB) \cite{harriott1986integrated}, Helium Ion Microscopy (HIM) \cite{peng2020source} and several other similar techniques are used. All of these techniques use active imaging where charge carriers are bombarded on the sample to generate a secondary response. SEM, being widely available and reasonably affordable, is the most common tool used for imaging ICs. TEMs are seldom used owing to their stringent sample preparation requirements as compared to SEMs \cite{lau2006non}. The typical cost associated with renting an SEM is around 100 euros for an hour \cite{courbon2015semba}.

\begin{figure*}
    \centering
    \includegraphics[width=\linewidth]{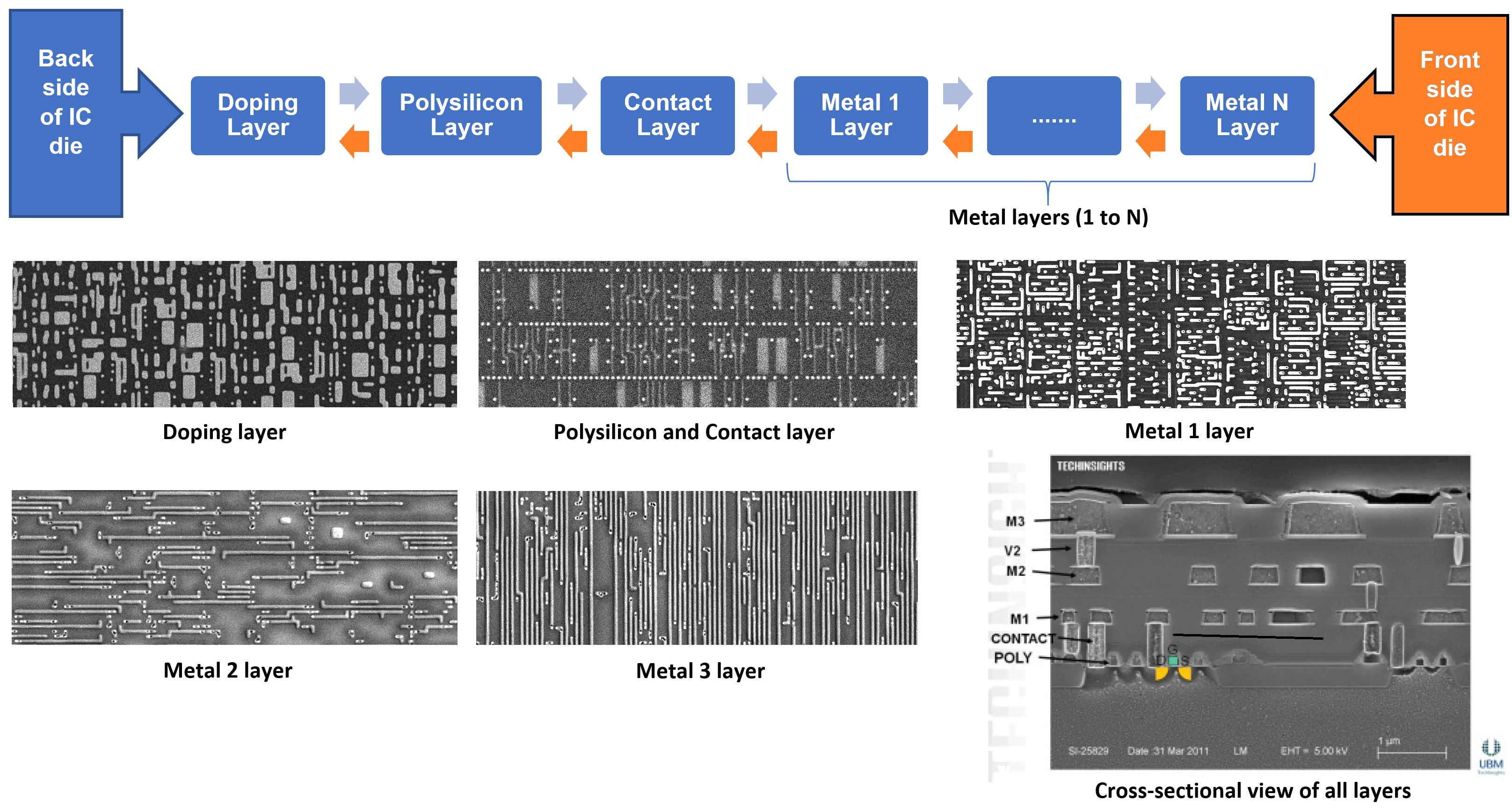}
    \caption{The sequence of the layers in an IC along with their cross-sectional view \cite{quadir2016survey, techinsightCircuit}}
    \label{fig:icLayers&crossSection}
\end{figure*}

\begin{figure}
    \centering
    \includegraphics[width=0.8\columnwidth]{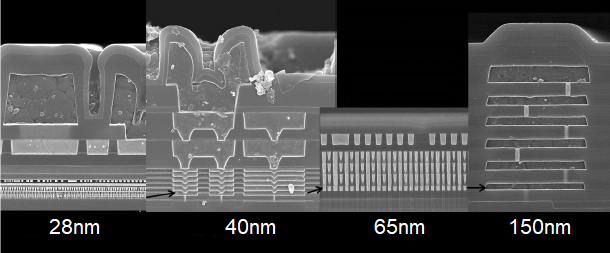}
    \caption{Cross-sectional view of ICs in various node technologies \cite{lippmann2019integrated}}
    \label{fig:cross_sec_all_node_tech}
\end{figure}

Destructive imaging is one of the most commonly used approaches for RE. It requires the sample to be mounted onto a plate, and layers of specific thickness are removed at a time. This process is called delayering. This is an iterative process with its ultimate purpose of imaging various IC layers (see Figure \ref{fig:icLayers&crossSection}). The depth of delayering and number of layers in the IC obtained using a cross-sectional image of the IC. \newversion{The cross-sectional view is obtained by vertically slicing the IC die using a diamond saw followed by mechanical polishing \cite{crosssec_howto}. 
After this, the IC die cannot be used for further processing.
Note that when performing destructive deprocessing for RE, the layer thickness should be known apriori through, e.g., access to the IC design information. }
A cross-sectional view of some contemporary ICs are shown in Figure \ref{fig:cross_sec_all_node_tech}. The delayering process uses wet/dry chemistry to etch away material from the IC surface to expose each layer. This process is also done mechanically using a multi-axial Computed Numeric Control (CNC) mill. The acid for etching the surface is selected depending on the material composition of the IC. A detailed description of the delayering process, including various etch chemistry, can be found in literature \cite{koh2011semiconductor, courbon2015semba, kison2019advanced, kimura2020decomposition}. 

At the slightly increased cost and deprocessing time overhead, a FIB can be used to delayer the sample more accurately. A FIB uses a focused ion beam to remove materials slowly with selections of gas chemistry \cite{wang2017probing, fyrbiak2017hardware}. FIB is generally a preferred option for ICs using contemporary node technology because of their smaller feature sizes and fragile nature. It has been recommended that at least one die be provided for each layer in the IC \footnote{Chipworks, currently TechInsights. Link: https://www.techinsights.com/}. In the case that the number of samples is not adequate, it would be safer to opt for a non-destructive approach. 
Moreover, to improve the time-complexity of RE, in situations where the emphasis is not on extracting all the elements in the IC, but on a certain aspect of the IC, a Region of Interest (ROI) -based approach is taken. RE, in this case, is only performed on the ROI of the IC. This significantly reduces the effort required for RE.

\begin{figure}[t]
    \centering
    \includegraphics[width=0.75\columnwidth]{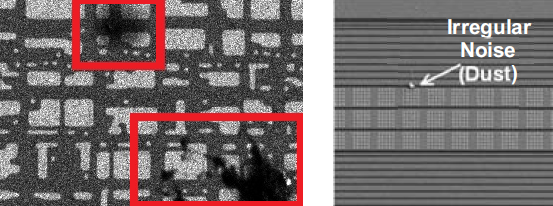}
    \caption{Effects of inadequate care during the sample preparation process. Etch residue leftovers (left) and dust particle on the die (right) \cite{hong2018deep}}
    \label{fig:etch_dust}
\end{figure}
\begin{figure}[t]
    \centering
    \includegraphics[width=0.35\columnwidth]{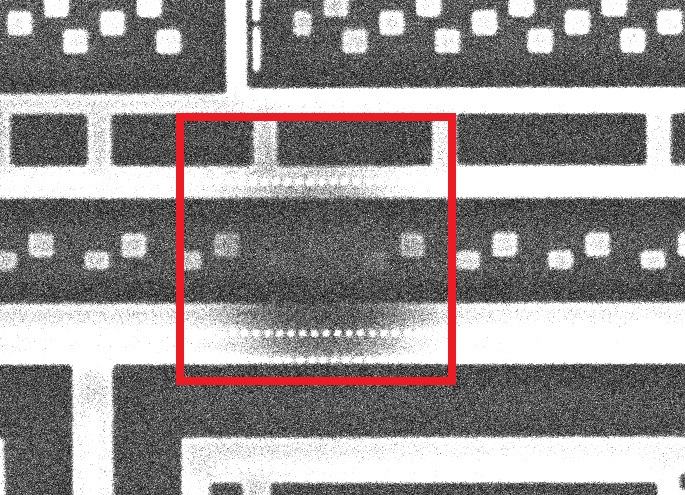}
    \caption{Effect of uneven delayering. This maybe a result of improper mounting of the sample or inadequate polishing of the die surface}
    \label{fig:uneven_delayer}
\end{figure}

Because of the importance of image acquisition for RE, clear precautions should be taken for sample preparation. For instance, as shown in Figure \ref{fig:etch_dust}, the etching process may leave some residue on the sample, or they may be dust particles on the imaging surface of the die. These should be cleaned thoroughly. Similar issues arise with milling and over-etching. Polishing must be carried to ensure a smooth sample surface for imaging. Earlier works in RE also provide cautionary reports on features from multiple layers being visible at the same time due to uneven delayering and improper mounting of the sample (see Figure \ref{fig:uneven_delayer}). Delayering should be performed parallel to the die base for optimal results. For technology nodes smaller than 10 nm, difficulties in achieving proper mechanical polishing and the tendency for surface damage have been reported in a case study \cite{sikul2018sem}.   

\begin{figure}
    \centering
    \includegraphics[width=0.8\columnwidth]{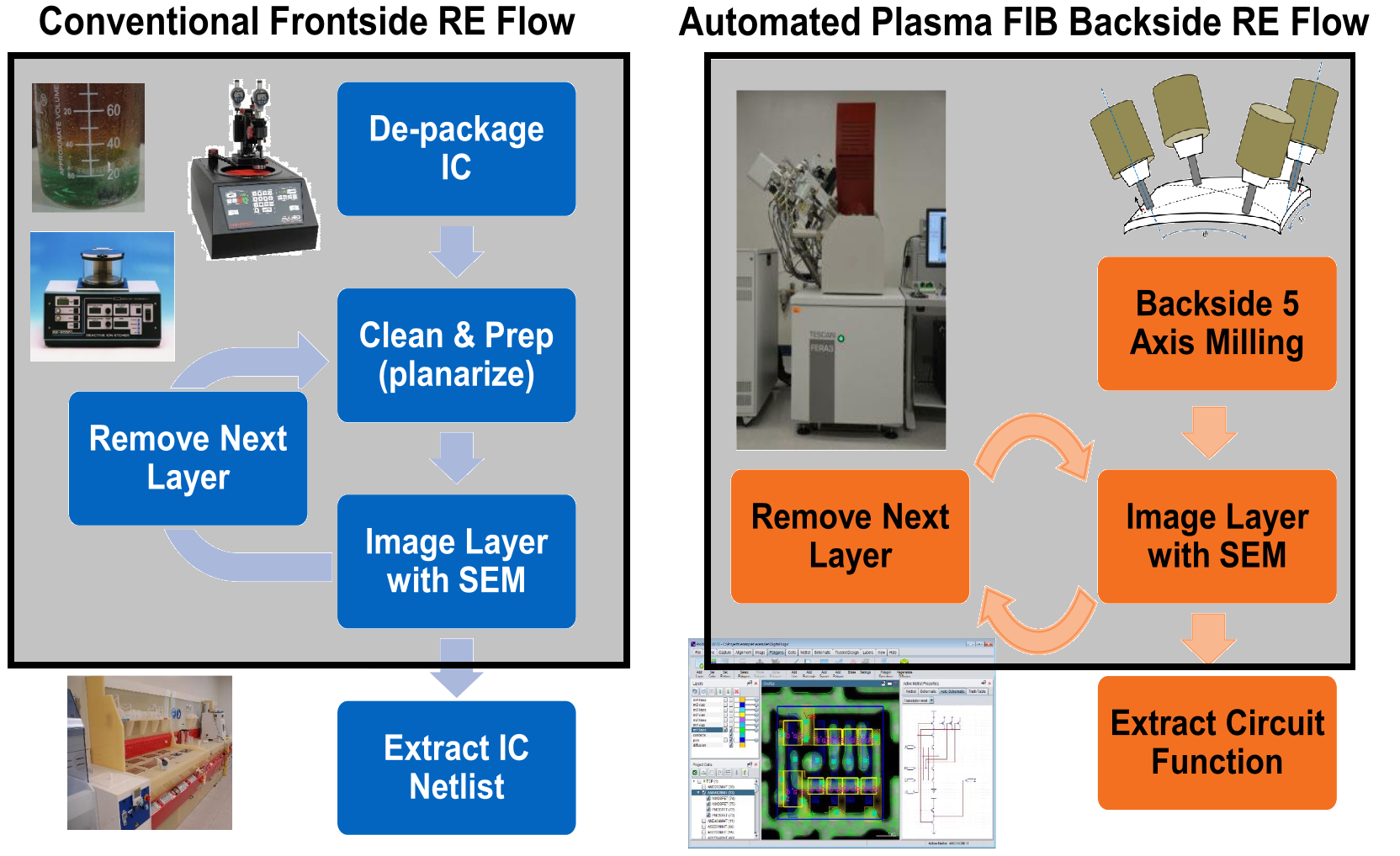}
    \caption{Front-side vs. Back-side deprocessing workflow \cite{principe2017steps}}
    \label{fig:front_back}
\end{figure}

\begin{figure}
    \centering
    \includegraphics[width=0.5\columnwidth]{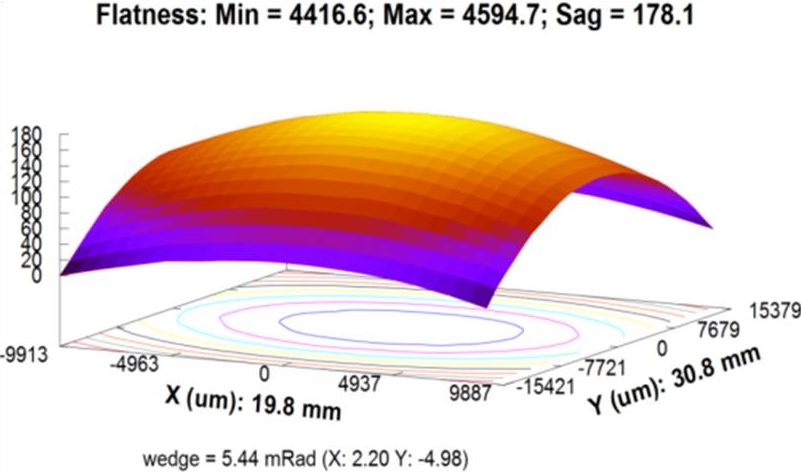}
    \caption{Warpage in the silicon die due to the accumulation of mechanical stress. Delayering on a warped die can cause uneven delayering and pin-cushion effects \cite{principe2017steps}}
    \label{fig:die_warp}
\end{figure}

In a typical RE process, the delayering operation is performed from the frontside of the IC. The frontside of the IC begins with the lowest metal layer in the structure. Being a well known approach, there are counter RE measures installed on ICs to prevent frontside RE \cite{castro2016frontside}. Although this countermeasure can be bypassed by a skilled operator using existing delayering methods, critical damages to the IC under deprocessing could be caused \cite{sikul2018sem}. An alternative to frontside RE is the backside RE. At present, there are no known countermeasures to backside RE \cite{kison2019advanced}. Initially developed as an approach for live-probing an IC under test for fault analysis, backside deprocessing has also been utilized for IC RE. A complete workflow for both approaches is shown in Figure \ref{fig:front_back}. Apart from the obvious advantage of safer sample deprocessing, the backside approach can also be fully automated \cite{principe2017steps}. A major drawback of backside deprocessing is the die warpage. With most of the bulk silicon substrate at the backside of the IC, deprocessing results in an increased mechanical stress on the die, causing it to warp. This phenomenon is shown in Figure \ref{fig:die_warp}. The sample preparation and delayering process typically take half a day to complete \cite{courbon2016reverse}. A detailed breakdown of the sample preparation time frame from a case study is shown in Table \ref{tab:time_deprocess}. 
\newversion{The FIB was introduced earlier for safer deprocessing. Although it can be used for both frontside and backside deprocessing, it is predominantly used for backside deprocessing. As shown in Figure \ref{fig:cross_sec_all_node_tech}, in typical ICs, the frontside is packed with thick metal layers. Being a process designed for precise removal of materials, FIB editing takes a considerable amount of time to get through the thick metal layers. For backside deprocessing, the bulk of the silicon substrate can be removed using simple mechanical polishing after the thickness of the silicon substrate is identified using a cross-sectional view. The last few nanometers of the silicon substrate can be handled using FIB for precise and time-efficient deprocessing. }

\begin{table}[t]
    \caption{Breakdown of sample preparation on different layers for a 130nm node technology IC (Die Area: 558,000 $\mu$m\textsuperscript{2} \cite{kimura2020decomposition}}
    \label{tab:time_deprocess}
    \centering
    \scriptsize
    \begin{tabular}{|l|c|c|c|c|}
    \hline
        
        Layer Name $\rightarrow$ & Metal 2 & Metal 1 & Polysilicon & Active \\
         \hline
        Time (mins) $\rightarrow$ & 232 & 199 & 26 & 42 \\
         \hline
    \end{tabular}
\end{table}

\subsubsection{Imaging}\label{subsubsec:imaging}

\begin{figure*}[t]
\begin{subfigure}[t]{1\textwidth}
\centering
\includegraphics[width=0.8\columnwidth]{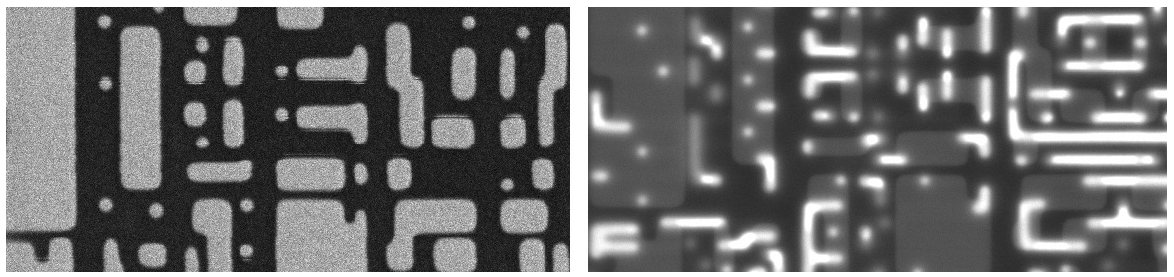}
\caption{Excitation voltages: the image acquired at 5 kV (left), which shows the surface-level features for the doping layer, and the image acquired at 15 kV (right) showing the same features of the doping layer along with the features of the contact/metal layers, present below the surface. }
\label{SEM_potential}
\end{subfigure}

\begin{subfigure}[t]{1\textwidth}
\centering
\includegraphics[width=0.8\columnwidth]{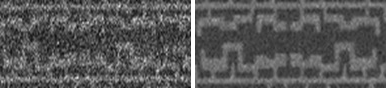}
\caption{Dwelling time: 3.2 $\mu$s/pixel (left) and 32 $\mu$s/pixel (right) \cite{vashistha2018detecting}.}
\label{SEM_dwelltime}
\end{subfigure}

\begin{subfigure}[t]{1\textwidth}
\centering
\includegraphics[width=0.8\columnwidth]{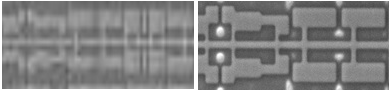}
\caption{Magnification: 500~$\mu$m (left) and 20~$\mu$m (right) \cite{vashistha2018detecting}. }
\label{SEM_mag}
\end{subfigure}

\begin{subfigure}[t]{1\textwidth}
\centering
\includegraphics[width=0.8\columnwidth]{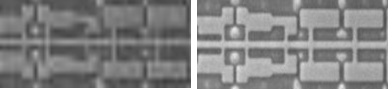}
\caption{Resolution: 512 x 512 (left) and 2048 x 2048 (right) \cite{vashistha2018detecting}. }
\label{SEM_res}
\end{subfigure}
\caption{The impact of changing the imaging parameters.}
\label{fig:SEM_total}
\end{figure*}

Destructive RE is the most common approach in the RE community. Being an iterative procedure, as illustrated in Figure \ref{fig:re_flow}, delayering and imaging are done in an alternating fashion to acquire images of all layers in the IC. 
As mentioned earlier, SEM is the most common imaging modality used in RE, owing to its simplicity and ease of access. Consequently, we will be focusing on SEM in this tutorial. In SEM imaging, the following parameters are commonly fine-tuned depending on the features that need to be extracted:

\begin{itemize}
\item \textit{Excitation voltage:} The excitation voltage of the electrons controls the depth of penetration into the sample. A higher excitation voltage can show structures that are hidden below the visible surface (see Figure \ref{SEM_potential}).
\item \textit{Dwelling time:} This refers to the time that the scanning beam takes to measure the intensity value of a single pixel in the image. A longer dwelling time would give a better estimation of the true intensity of the value at the given position (see Figure \ref{SEM_dwelltime}).
\item \textit{Field-of-View:} Commonly referred to as magnification. It refers to the size ratio between an object and its scaled projection. A high level of magnification enables us to take images of small features that cannot be seen at low magnification (see Figure \ref{SEM_mag}). 
\item \textit{Resolution:} This parameter refers to the number of pixels in the image. Higher pixel count produces better images (see Figure \ref{SEM_res}). 
\end{itemize}

The time cost of taking images for a 130~nm IC is reported in Table~\ref{tab:SEM_time} \cite{vashistha2018detecting}. 
It can be observed that full-blown RE for a single layer, with high-quality image acquisition settings, takes over 30 days to complete. 

\begin{table}[t]
\caption{Time table of SEM imaging for an IC with the following characteristics: technology node: 130nm, and size: 1.5~mm x 1.5~mm \cite{vashistha2018detecting}}
\label{tab:SEM_time}
\scriptsize
\begin{center}
\begin{tabular}{@{}|c|c|c|c|@{}}
\hline
Scanning Speed & \begin{tabular}[c]{@{}c@{}}Field of View\\  Resolution\end{tabular} & 500um x 500um & 20um x 20um \\
\hline
1 usec/pixel & 512x512 & 9 sec & 1 hr 33 min \\ 
\hline
1 usec/pixel & 1024x1024 & 18 sec & 3 hr 7 min \\
\hline
1 usec/pixel & 2048x2048 & 54 sec & 9 hr 22 min\\
\hline
10 usec/pixel & 512x512 & 45 sec & 7 hr 48 min\\ 
\hline
10 usec/pixel & 1024x1024 & 3 min 18 sec & 1 d 10 hr\\ 
\hline
10 usec/pixel & 2048x2048 & 6 min 25 sec & 5 d 12 hr\\ 
\hline
32 usec/pixel & 512x512 & 1 min 30 sec & 15 hr 0 sec \\ 
\hline
32 usec/pixel & 1024x1024 & 6 min 30 sec & 1 d 21 hr\\ 
\hline
32 usec/pixel & 2048x2048 & 24 min 0 sec & 11 d 1 hr\\ 
\hline
100 usec/pixel & 512x512 & 4 min 48 sec & 2 d 2 hr \\ 
\hline
100 usec/pixel & 1024x1024 & 18 min 54 sec & 8 d 4 hr\\ 
\hline
100 usec/pixel & 2048x2048 & 1 hr 11 min & 30 d 20 hr\\ 
\hline
\end{tabular}
\end{center}
\end{table}

A rule of thumb in image acquisition is to use higher magnification, resolution, and dwelling time; however, these choices significantly affect imaging times. Imaging potential is typically kept low ($\leq$5kV) to prevent interaction with underlying layers. For more control, a few more parameters can be adjusted \cite{lippmann2019integrated}. The goal for imaging is to effectively resolve the features in the IC. For instance, in a case study, each pixel in the image corresponded to 10 nm in the actual IC \cite{lippmann2019integrated}. If the smallest feature in the IC is smaller than the resolution capability of the SEM, the feature cannot be resolved effectively. With studies suggesting features spanning a few pixels in the image being a cause of concern for the hardware assurance community, a higher magnification is usually warranted. Approaches that can automatically, dynamically control the SEM magnification based on the smallest feature size in the current Field-of-View have been proposed to optimize the magnification selections for real-world applications \cite{wilson2019first}.

\begin{figure}
    \centering
    \includegraphics[width=0.7\columnwidth]{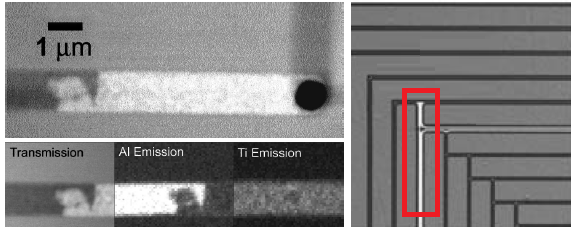}
    \caption{Electron migration (left) due to accelerated life cycle conditions causing variations in intensities measured for the same material \cite{frigo2002submicron}. Charging effects (right) exhibited in insulating materials under SEM observation \cite{hong2018deep}}
    \label{fig:emig_charging}
\end{figure}

There are some inherent drawbacks of utilizing electron microscopy for imaging IC structures. Some of them result from the physics behind the imaging modality, whereas other ones are inherent to the semiconductor structure and associated deprocessing. They are discussed below in detail:

\begin{itemize}
\item \textit{Manufacturing process variations:} Due to the high accuracy of the imaging modality, any small variation in the manufacturing process would cause changes in the acquired image. The degree of influence of these variations depend on the precision/tolerance of the manufacturing process and the resolution of the imaging modality. These variations are utilized in hardware assurance techniques such as PUFs. Being naturally nondeterministic processes that may not necessarily be parametric, modeling these variations using statistical models is not recommended \cite{sarkar2019automating}.
\item \textit{Topography of the material:} Areas with high roughness or edges between materials in the IC have larger escape areas for the secondary electrons \cite{harriott1986integrated, seiler1983secondary}. These cause intensity variations of the same material measured in different locations on the same IC die. This further emphasizes the need for polishing and planar surface on the deprocessed IC die. 
\item \textit{Oxidation:} Delayering exposes the metallic structures in the IC to the atmosphere \cite{sikul2018sem}. Oxidation of metallic surfaces causes fluctuations in the responses obtained for the same material at various points in the IC. 
\item \textit{Electron-migration:} If the IC has been under use, there are chances of having electron migration and changes in the physical structure of the material  \cite{frigo2002submicron}. This type of defect is usually found in metal interconnects, through which high-density currents flow (see Figure \ref{fig:emig_charging}). This issue is predominantly studied in device failure analysis in estimating the lifetime of an IC \cite{malone1997electromigration}.
\item \textit{Conductivity:} Insulating materials may charge positively and suppress the secondary electrons \cite{harriott1986integrated}. This leads to localized pockets of bright and dark regions in the image (see Figure \ref{fig:emig_charging}). This issue can be offset by depositing thin layers of conductive materials such as carbon or platinum on the IC die surface \cite{quijada2018large, koh2011semiconductor}. Moreover, if the material is considerably thin, the electrons can pass through so that the sensors cannot detect them \cite{salzer2012two}. It should also be noted that when using an active mode of imaging, radiation exposure over a long duration at a specific location causes a contamination layer to build up over the exposed region,  preventing further emission of electrons for effective imaging \cite{seiler1983secondary}.  
\end{itemize}

\begin{figure}
    \centering
    \includegraphics[width=\columnwidth]{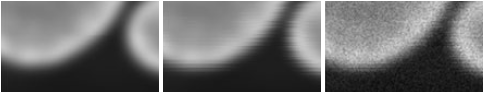}
    \caption{Practical limitations of electron microscopy techniques. The raw image is shown on left and the middle one is corrupted with drift and vibration. Rightmost image is corrupted with all three noise sources (drift, vibration, and radiation dose). Context: Gold particles on carbon surface \cite{cizmar2008simulated}. }
    \label{fig:noise_imaging}
\end{figure}

Several other issues that are predominantly associated with electron microscopy are illustrated in Figure \ref{fig:noise_imaging} as an exemplary case.
\begin{itemize}
    \item \textit{Drift}: It is present in every EM modality and cannot be fully accounted for in every case. It occurs mainly due to random fluctuations in the scanning beam and mechanical creep induced by the staging platform of the SEM. Drift is more pronounced at a lower magnification that covers a larger Field-of-View \cite{cizmar2008simulated}. 
    \item \textit{External Environment}: Vibrations, even those that are barely perceptible, along with thermal expansions in the sample caused by slight temperature fluctuations in the environment, can significantly affect image quality \cite{cizmar2008simulated}.    
    \item \textit{Radiation dose}: This specific aspect of imaging is controlled by the dwelling time of the beam over a given pixel. In simple terms, longer dwelling times result in better images. With significant development in SEM imaging, especially in scan generators, it is possible to acquire significantly better images at lower dwelling times. However, there is a persisting interest in the community for low-dose imaging to reduce sample damage \cite{giannatou2019deep}. There have been several attempts to model the noise introduced by lower dose imaging in the microscopy community, but the influence of noise in recovering feature in semiconductor RE has not been studied in detail. Currently, there is a consensus agrees on modeling the noise as a Poisson-Gaussian process \cite{kockentiedt2013poisson, pawley2007biological, frank2005noise, sakakibara2019impact, sim2004effect}.
\end{itemize}
With regard to the last point, the noise from the electronic components of the SEM (e.g., amplifiers and scan generators) has been modeled as Additive White Gaussian Noise (AWGN), but its influence was found to be negligible as compared to that of emission noise \cite{batten2000autofocusing, timischl2012statistical}. It has been suggested that the noise introduced by the imaging modalities can be used to hide hardware Trojans \cite{sarkar2019automating}. This is the motivation behind studying noise in SEM and acquiring good quality images for hardware assurance applications. The RE workflow has not been optimized for repeatability. Hence, there will always be unforeseen circumstances during the RE process that would require human interaction to resolve.

\begin{figure}
    \centering
    \includegraphics[width=0.65\columnwidth]{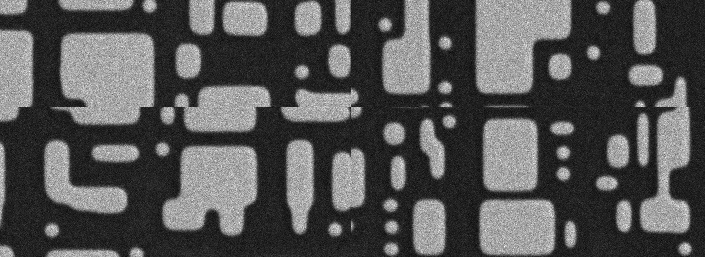}
    \caption{Stitching error in the active layer due to improper image registration}
    \label{fig:stitch_error}
\end{figure}

In addition to the above issue, when employing SEM, caution regarding the following problems should be used. 

\vspace{5pt}\noindent\textbf{Stitching process: }
In most cases, the Field-of-View provided by SEM does not cover the entire ROI. This typically results in multiple images to be collected and stitched together to form a complete image. In typical applications, the stitching process involves taking two images with a certain overlap. The degree of overlap is decided by the operator and remains fixed for the entire image acquisition phase. Similarity measures like normalized cross-correlation \cite{lippmann2019integrated} or features like Scale Invariant Feature Transforms (SIFT) \cite{6808796} are used to find the position with the highest correlation. Then the two images are merged by blending the overlapping region or by just using one of the overlapping regions and discarding the other. Stitching is usually an error-prone process, where the error can usually be resolved by a subject-matter expert. There have been attempts at using more advanced approaches for improving stitching accuracy. 

Nevertheless, stitching for contemporary nanoscale node technologies is even more challenging. 
Specifically, in such modern ICs, especially in metal layers, the features are very much similar and repetitive, causing significant corruption in the stitching process. Correlation analysis and Fourier domain analysis have been conducted in some studies to demonstrate this issue \cite{wilson2019novel, zhang2019fast}. An ingenious approach has been suggested utilizing leftover imperfections from the delayering process to stitch images with similar features \cite{zhang2019fast}. These imperfections in deprocessing are usually very unique and unlike inherent features in the IC under study. This work also discusses the issues with stitching images in all directions on the same plane and the importance of optimizations for ensuring the best fit among neighboring images in the plane. An example of a stitching error is shown in Figure \ref{fig:stitch_error}.     

\vspace{5pt}\noindent\textbf{Alignment process: }The images taken from layers are stacked on top of each other. The alignment is done using correlation matching, typically involving vertical interconnects (VIA). Design rule checks and manual operator intervention validate the vertical alignment of the image stack. There are several software suites that are available to ensure seamless alignment and viewing of images such as Chipwork's IC Browser \cite{torrance2009state}, AIRE \cite{bowman2018image}, Olyvia \cite{koh2011semiconductor} and Hugin  \cite{courbon2015semba}.

\begin{figure}
    \centering
    \includegraphics[width=\textwidth]{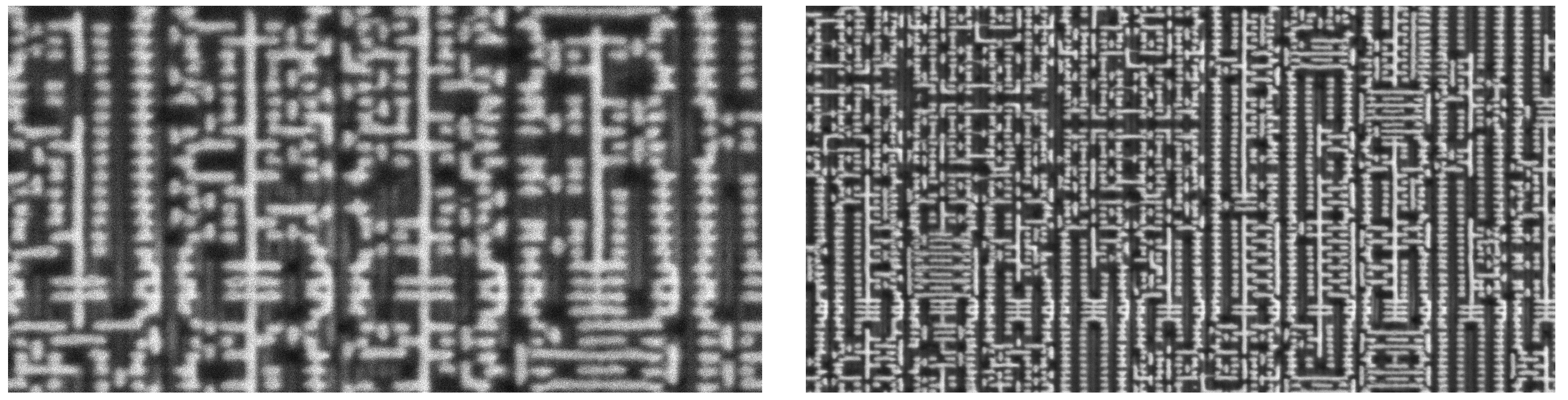}
    \caption{SEM images acquired from a FinFET IC utilizing 14nm node technology. Field-of-View has been varied in both images to display contrast in feature sizes. Left: 5$\mu$m, Right: 15$\mu$m.}
    \label{fig:finfet14nm}
\end{figure}

\vspace{5pt}\noindent\textbf{Emerging technologies: }Although there is no exemplary work done for RE on contemporary technologies such as FinFET, some key insight can be obtained from challenges faced by the Fault Analysis community \cite{orji2018metrology}. Some dielectric materials have low contrast with the silicon substrate making them harder to identify. A similar situation is shown in Figure \ref{fig:finfet14nm} with the contrast between the contacts and the metal layers significantly reduced as compared to previous node technologies. This makes it considerably harder to visually analyze the image and hinders contrast-based image analysis techniques. It has also been suggested that samples sustain more damage under typical radiation dose as compared to older node technologies. There are studies that combine simulation-based methods with SEM imaging for higher quality images at lower radiation dose. These hybrid methods are shown to be effective for recent node technologies. This further emphasizes the need for better noise models for noise-free image acquisition at lower dwelling times. Finally, there are challenges associated with deprocessing layers with very fine features. With a higher level of integration in ICs such as 3DVLSI, a full-scale RE workflow would be a daunting task without significant automation in imaging, deprocessing, and error resolution. The challenges related to performing RE on post-CMOS technology nodes are yet to be explored.

\vspace{5pt}\noindent\textbf{A note on the positive side of SEM: }
A limitation of data-driven approaches is the unavailability of good quality data, as stated earlier. 
To address this, the electron microscopy community introduced an SEM image simulator called ARTIMAGEN \cite{cizmar2008simulated, cizmar2009optimization}. 
However, in recent years, there have been attempts to generate synthetic data for evaluating data-driven approaches. This simulator can generate images with varying influences of drift, vibration, thermal expansion, and noise profiles (Gaussian/Poisson). However, the selection of materials, noise profiles, and shape contours are limited. The fault analysis community extensively uses simulated images for benchmarking Line Edge/Width Roughness (LER/LWR) algorithms \cite{chaudhary2019line}. A deep learning approach has been recently suggested to generate synthetic SEM images based on layout data for mask optimization and virtual meteorology \cite{shao2020ic}. Although the data used for these approaches are not publicly available, it does provide a path towards integrating data-driven approaches for RE in the imaging phase.  

When it comes to applications of SEM in RE, there are some commercial solutions available. Notable devices include the Multi-SEM (mSEM) \cite{keller2014high} from Zeiss for using multiple scanning beams for faster high quality image acquisition and Chipscanner 150 \cite{lippmann2019integrated} from Raith for use in large-scale image acquisition with higher accuracy requirement on stitching. In contrast to the timing requirement shown in Table \ref{tab:SEM_time}, Chipscanner 150 has a reported scan time of 4.9 hours for imaging at 4000x4000 resolution with a 40$\mu$m field-of-view on a 1mm\textsuperscript{2} area \cite{lippmann2019integrated}. In the same work, this scan time was brought down to 0.7 hours with a better scan generator. Although convenient, these systems are also significantly more expensive than regular SEMs \cite{trampert2018should}. The computational resource complexity is also an attention-worthy aspect of RE. For instance, the storage complexity associated with acquiring images of the entire active layer of a 45nm node technology IC has been estimated to be over 22 gigabytes on regular SEMs \cite{courbon2019practical}. For high-performance imaging systems like the mSEM, this requirement is much higher \cite{zhang2019fast}.

\subsection{Problems Associated with Handling PCB Images}
As opposed to what has been explained for ICs in Section~\ref{sec:image_IC}, the imaging process performed on PCBs does not require decapsulation. 
However, components mounted on the PCB can be removed for easier access to the PCB markings and easier imaging for multi-layered PCBs. The PCB equivalent of decapsulation in ICs is referred to as de-soldering, or the removal of componentes on the PCB. The byproduct of this de-soldering is varying levels of collateral damage to the PCB, depending on the efficacy of the decapsulation process. This will have different impacts on later stages of RE, such as image acquisition, where the damages are reflected in the form of debris, physical distortion, or noise artifacts present in the digital image. Therefore, it is recommended to avoid de-soldering when possible, but if necessary, it should be done with the utmost care as to limit the damage to other key features (trace tracks, via conductive rings, text markings on PCB surface, etc.). 

The challenges and limitations associated with PCB RE overlap significantly with those presented earlier in IC RE scenarios. The main difference is that PCB RE has two focused aspects: external and internal RE. \emph{External} PCB RE deals with the information that one can observe at both surfaces, the top and bottom layers of a PCB. This information typically consists of the components of a design (passive elements, active elements, ICs, processors, etc.), their connections, silkscreen markings, and a variety of ports \cite{quadir2016survey}. External RE would usually suffice if the PCB has only two layers, but this is often not the case. More common, however, are PCBs manufactured with multiple layers, where the majority of them are \emph{internal} to the board and have structural and connectivity information not visible externally. In these cases, internal RE is necessary \cite{quadir2016survey}. 

Traditionally, internal RE has been a destructive process \cite{grand2014printed} similar to that of IC RE. The process involves delayering and imaging of a PCB layer-by-layer until a working functional physical sample no longer exists. The imaging component of this process is typically done optically by using a digital camera or a high-quality optical microscope, but the destructive nature of the delayering process introduces multiple potential sources of the noise that could impact the quality of the image. Some examples include broken traces, disconnected vias, or poor quality images, making feature extraction much more difficult for analysis. These features need to be imaged to the highest degree of detail possible to facilitate optimal performance in the RE process, which is usually bound by the quality of the image under analysis. 
Fortunately, recent progress toward non-destructive RE via X-Ray CT has pushed the current state-of-the-art RE methods \cite{asadizanjani2015non}.

\subsubsection{Imaging}
The main challenges for each PCB RE modality can be broadly categorized into how to handle the noise associated with the imaging modality used for data acquisition.
We expand on these issues for both the external and internal RE, as defined above. 

\vspace{5pt}\noindent\textbf{External RE: }The imaging modality used for data acquisition in external PCB RE is typically an optical microscope or a digital camera. Both are used to take images of a PCB at varying resolutions to enable the detection, classification, and analysis of the design information. Specifically, external RE uses these images to identify the components, connections, silkscreen markings, and different types of ports (high-speed serial/parallel, program/debug, display) present on the topside as well as bottomside of a PCB \cite{torrance2009state, grand2014printed}. Among all imaging modalities, the illumination variance is the most prominent noise source. In some cases, imaging an entire PCB board requires stitching, which results in multiple regions of the entire board with varying illuminance. This variation may cause differences in the appearance of even the same sample; therefore, drastically impacting the effectiveness of image analysis algorithms and the inspection results. Hence, a dedicated optical imaging station that provides consistent illumination across the entire imaged board is recommended. The dedicated imaging station should consist of a constant light source that is well distributed across an area. For example, a hanging lamp in combination with a high-resolution DSLR camera can be used to equip a larger station and an optical microscope station used for small to medium samples. 

\begin{figure}[t]
    \centering
    \includegraphics[width=0.5\columnwidth]{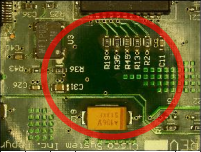}
    \caption{Optically imaged PCB section highlighting illuminance non-uniformity}
    \label{fig:lightingNonUniform}
\end{figure}

Depending on the magnification of the microscope or camera used for data acquisition, the image patches obtained vary in size and the amount of information included in them. While an increase in the magnification obtains more detailed features for extraction, the illumination noise is also amplified (see Figure \ref{fig:lightingNonUniform}), causing an information loss. For instance, using low magnification results in a larger image view, but we may lose small features (e.g., characters on resistor) due to the reflection. This could make the detection of Trojans, i.e., maliciously inserted/replaced components, more challenging. Although more features per image lead to more details, more image patches should be stitched together to complete the whole image, and thus, regions with the various illuminance are involved in the fully stitched sample. 

On the contrary, when the image magnification increases to obtain more details, some large components on a PCB are separated into different patches. Moreover, since many of the existing image analysis algorithms for segmentation, detection, and classification are heavily parameter-dependent or pixel intensity sensitive~\cite{aggarwal2006line,Kalviainen1995ProbabilisticComparisons,kass1988snakes,schubert2017dbscan,maini2009study,dalal2005histograms}, a holistic solution for PCB RE that is more generalizable by minimizing parameter tuning is needed.

\vspace{5pt}\noindent\textbf{Internal RE: }If the PCB under RE has only one to two layers, the challenges encountered by the expert would be solely limited to those discussed above. However, it is more likely that modern PCBs are multilayered, where chips are connected to each other on the top, bottom, and through internal layers. Therefore, for multilayered PCBs, internal RE is required to complete the RE process. When discussing internal RE, there are two predominant methods: destructive and non-destructive RE. As explained earlier, the destructive approach is obsolete by contemporary standards and will not be considered further.  

\begin{figure}
    \centering
    \includegraphics[width=0.5\columnwidth]{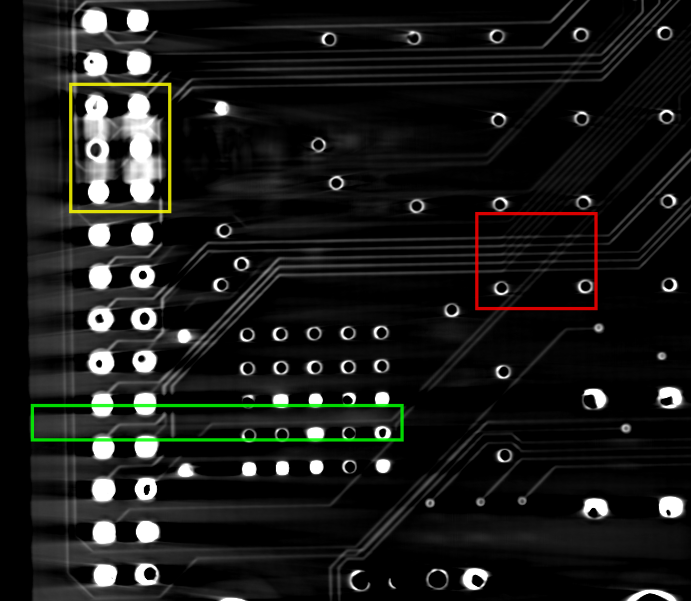}
    \caption{Noise in X-Ray CT PCB:(red square on the right side) Neighboring Layers Aliasing (green square on the bottom left side) Blur Artifacts (yellow square on the top left side) High Impedance Material Artifacts}
    \label{fig:xrayNoise}
\end{figure}

While non-destructive RE via X-Ray CT does reduce the amount of physical damage to a PCB caused by the delayering process, there are other challenges still to be faced, which impact the quality of the RE. These noise sources generated by the X-Ray process are outlined below.

\begin{itemize}
    \item \textit{Blur artifacts:} During the X-Ray CT process, the sample is rotated inside the X-Ray chamber by 360\degree at a slight tilt in the Cartesian plane to maximize the amount of information received from the X-Ray particles passing through the sample to the receiver in the chamber. This tilt, along with the rotation during the acquisition process, results in noise artifacts in the reconstructed 3D stack in the form of blurred image regions.
    \item \textit{High impedance materials:} Typically, PCBs are manufactured with the majority of their parts made of silicone-based materials. However, if the PCB is populated with components, soldering is used to ensure the connectivity of the components throughout the entire board. This solder acts as a high impedance material in the presence of X-Ray particles, reducing the effectiveness of their passing-through property. Therefore, it creates noise artifacts in a populated sample or a sample with components being removed, but with remaining solder residue. 
    \item \textit{Aliasing between neighboring layers:} An X-Ray CT-based PCB model is a 3D stack consisting of 2D image slices. Each layer of a multi-layer PCB consists of the trace and via information that may differ from those being close to that layer. Depending on the alignment of the board in the X-ray chamber, the resolution, and the X-Ray parameters chosen, there may be slices at the fringes of neighboring layers, where the information of the layers overlaps, similar to aliasing in signal processing. 
    \item \textit{Beam hardening:} While an X-Ray beam passes through an object, the mean energy of the beam increases as the low energy photons are attenuated \cite{boas2012ct}. Therefore, the lower energy part of the X-ray beam is removed from its energy spectrum, and the beam is considered to become ``harder". Due to this X-ray beam hardening, streaks or dark bands appear at the center of the object, compared to the edge of the object in the X-ray image. For this purpose, pre/post-filtering the X-Ray beam by using metallic materials, e.g., aluminum and copper, is applied to eliminate the low energy photons in the beam and maintain uniform average energy during the X-ray imaging. 
    \item \textit{Ring artifact:} In general, miscalibrated or defective detectors and elements create a bright or dark ring close to the isocenter of the scan. This can often be fixed by recalibrating the detector.   
\end{itemize}

The blur caused by the X-Ray is seen as the streaks in the image (see the green square on the bottom left side of Figure \ref{fig:xrayNoise}). The bright circular regions are where the solder has impacted the X-Ray process, distorting the via features slightly, as illustrated in the yellow part on the top left side of the figure. The areas, where the trace information is intersecting each other, show aliasing that occurs between the neighboring layers (see the red square on the right side of Figure \ref{fig:xrayNoise}). While these are the main sources of the noise seen during the X-Ray process conducted on a PCB, their effects are compounded when taking the reconstruction process for X-Ray CT into account. In particular, at each slice of the reconstructed PCB, crafted by using X-Ray CT, the noise sources have a varying degree of impact since they represent the varying depths of the board and the depth of the X-Ray particles passing through the samples. Therefore, it is necessary for image analysis-related solutions to not only account for the variance seen within a single design, but also across multiple designs. 
Nonetheless, the impact of the noise could be reduced by adjusting parameters discussed below. 

\begin{itemize}
\item \textit{Tube voltage:} This parameter adjusts the peak energy of the X-Ray beam (i.e., raises the average energy of the photons). The choice of the tube voltage affects the image contrast in the scanning process. An increase in this voltage leads to a lower contrast in the images. 
\item \textit{Tube current-exposure time product:} This refers to the number of photons produced per unit time. Random, thin, bright and dark streaks are considered as noise that may appear in the images due to the low photon counts. 
\item \textit{Resolution:} The resolution of the image is defined as the pixel size selected during X-Ray image acquisition, where the pixel size can be identified as the limiting factor for spatial resolution.
\item \textit{Filtration:} Filters are used to reduce the beam hardening effects in the X-Ray beam. As the low-energy photons do not penetrate through the object, filtration improves the quality of the beam.
\end{itemize}

\newversion{For the sake of comparison, Table \ref{tab:res_limit} provides the resolution limits of the imaging modalities and associated processes discussed in this section.}

\begin{table}[t]
    \caption{Resolution limits of imaging modalities and associated processes discussed in this tutorial}
    \label{tab:res_limit}
    \centering
    \scriptsize
    \begin{tabular}{|l|c|c|c|c|}
    \hline
        
        Imaging Modality $\rightarrow$ & SEM & FIB & X-ray CT & Optical (DSLR) \\
         \hline
        Resolution limits $\rightarrow$ & 2\textit{nm} & 15\textit{nm} & 1-2\textit{$\mu$m} for ICs/ 10-50\textit{$\mu$m} for PCBs/ 50-100\textit{nm} for wafer-level samples & 85\textit{$\mu$m} \\
         \hline
    \end{tabular}
\end{table}
\section{Addressing Challenges Associated with Machine Learning and Image Analysis for RE} \label{sec:section3}

In the context of natural scenes, image processing plays the role of enhancing the image to the point of being discernible and pleasing to the viewer. The fine-tuning of different image parameters, such as the contrast and intensity, is considered as an art rather than an application of a set of predefined algorithms. However, with the advent of machine learning and the higher likelihood of the image being delivered to a computer than a human, the adjustment of the imaging parameters must be performed regarding the application for a particular domain and nature of the problem being addressed. For instance, a camera placed on an assembly line in a manufacturing facility might only require to examine the presence of an object rather than its color or shape. Modifying the image to be visually pleasing is neither required nor recommended in this case. 

Optimizing images for a certain purpose requires in-depth knowledge of the domain and application, posing a significant challenge to the application of image processing in electronics, e.g., images taken from ICs or PCBs. It is reasonable to say that almost any data-driven approach, especially machine learning and image analysis, has its performance limited to the quality of the data provided to it. For instance, in the case of IC RE, the importance should be placed on electrical connectivity rather than visual quality. Assessing data quality using existing measures such as Signal-to-Noise (SNR) ratio does not provide insight into the true quality of the sample. Hence, invalidating the insight obtained using such samples.  

Along with the extensive application of image processing in natural scene-related scenarios, the factors affecting the quality of images, such as motion blur, sensor noise, and uneven lighting, are well known and studied. This in-depth understanding enables the development of image processing algorithms that can suppress the effect of noise sources, as mentioned above, and produce high-quality images. However, this does not hold true for imaging modalities used for acquiring images of ICs and/or PCBs. Hence, the wealth of knowledge compiled over time in the field of image processing cannot be effectively utilized for RE and hardware assurance.

\begin{figure}
    \centering
    \includegraphics[width=\columnwidth]{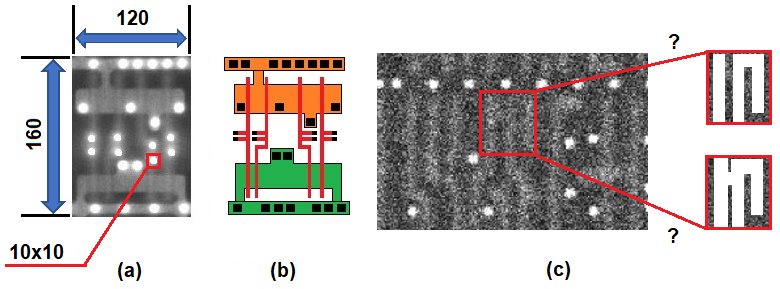}
    \caption{Difficulties with the handling of IC images: (a) an SEM image of a logic cell showing the size of features in pixels, (b) the software-based structure of the cell shown in (a), and (c) an example illustrating the ambiguity in the feature resolution in the presence of the noise.}
    \label{cell_size}
\end{figure}

Moreover, in typical natural scene images, the amount of information extracted is rarely dependent on the value of a few pixels, but on the entire image or a large section of the image. On the other hand, by increasing the level of integration that puts together a higher number of transistors into a limited space, the size of the features in an IC image usually expands to only a few pixels (see Figure \ref{cell_size} (a, b)). An analogous task for PCB RE is detecting vias and traces in a densely populated image where these features and structures can be only a few pixels long or wide. Hence, depending on the intensity of noise affecting such features, there are situations, where a structure containing a few pixels cannot be categorized into a feature or a noise artifact (see Figure \ref{cell_size} (c)). In the context of hardware assurance, this ambiguity may result in miss classification of a Trojan. Similarly, PCB images used for RE are subject to this challenge, albeit in a different way, due to the different modalities the images are taken from. For instance, optical imaging can easily fail to detect an extra component acting as a Trojan, when the color of the motherboard and the components on it are both black. Furthermore, via detection through X-Ray CT imaging may not be helpful when vias are blurred by the presence of high impedance materials that attenuate X-rays. 

Challenges that can be resolved through machine learning have three factors in common: a coarse understanding of the problem, the availability of high quality labeled/unlabeled data, and understanding of the statistical features that can assist in quantifying/representing the problem. With hardware designers and manufacturers having design experiences that are specific to them, it could be hard to find features that can be generalized to all cases. Even in approaches that support automated feature discovery, such as artificial neural networks, a significant amount of labeled data is required. In many instances, obtaining such ground truth is either not possible (obsolescence) or impractical (pixel-level labeling for semantic segmentation) due to the time and resources required. With issues compounding on every basic step of the machine learning paradigm, a holistic approach that can effectively cover all RE workflow stages is extremely challenging.     

The inherent complexity of using ML for RE can be demonstrated with a simple example. The steps involved in designing an IC usually start with a high-level description of all the modules. These modules are synthesized with standard cell libraries from IC manufacturers. The design is then placed, routed, and etched onto a silicon wafer die \cite{fyrbiak2019constructive}. In machine learning terminology, this whole process is akin to performing a dimensionality reduction technique, such as Principal Component Analysis (PCA), on the raw data. More specifically, although the processed low dimensional data behaves in the same fashion as the raw data and can be considered equivalent, recovery of the original data from a low dimensional projected version is hard. 

The goal of most RE applications is broadly classified into three types: recovery of schematics, obtain insights into the design, and to detect anomalies in the design for hardware assurance purposes. In order to satisfy these goals, the constraints on applying data-driven methods will have to be relaxed and explored within individual modules in the RE workflow, where the strengths of image processing and machine learning can be effectively utilized. In the following sections, we will explore various approaches that can get the reader as close to the original design data as possible.

\subsection{Problems Associated with Applying ML for IC RE}
The applications of ML algorithms in IC RE mainly include two types of methods employed to perform annotation and extract the netlist\footnote{A netlist is a graph that shows relationships between various components. }, described in this section. 
\subsubsection{Annotation}

\begin{figure*}
    \centering
    \includegraphics[width=0.85\textwidth]{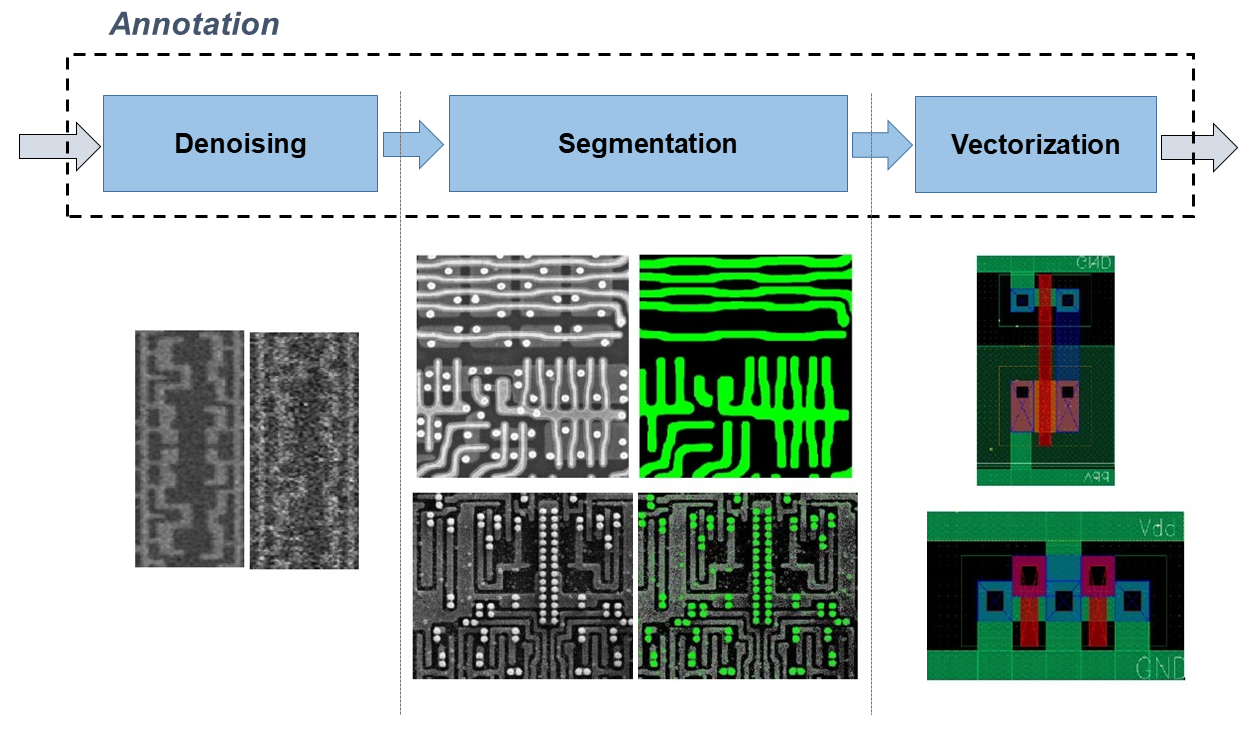}
    \caption{Expanded workflow for the Annotation block}
    \label{fig:annotation}
\end{figure*}

The annotation module expects image data from the previous step to be noise-free, free of deprocessing errors, well stitched, and aligned in all layers. However, in most cases, these are not fully satisfied. The annotation module can be divided into three sub-modules: denoising, segmentation, and vectorization. An overview of the annotation block can be seen in Figure \ref{fig:annotation}.

Prior to denoising the image, it is necessary to ensure that the image contents are not corrupted from deprocessing and stitching errors. Currently, there are no approaches to detect corrupted images. However, some insights can be obtained by looking into the forward manufacturing workflow. For instance, stitching errors may be corrected using design rule checks. Deprocessing errors are similar to faults in the IC die. There are quality control and design validation approaches that utilize image processing and pattern recognition to detect such faults. These approaches can be re-purposed for detecting deprocessing errors in the RE workflow \cite{nakagaki2010automatic, wang2007ic}. 

\vspace{5pt}\noindent\textbf{Denoising: }The denoising sub-module is responsible for ensuring that the noise component in the image is suppressed. There are several approaches currently employed in literature for processing noisy SEM images. They include spatial filtering approaches, including Gaussian, median, curvature, anisotropic diffusion, wavelet, adaptive wiener filter, and hysteresis smoothing \cite{masalskis2008reverse, 8311683, trindade2018segmentation, mazhari2016suppression}. Simple high-frequency filtering and deep learning-based denoising approaches have also been used on SEM images \cite{giannatou2019deep}. These techniques are mostly naive image processing techniques and do not take the semantics of structures in the image into account. Machine learning-based denoising approaches, such as image inpainting, super-resolution and dictionary-based sparse reconstruction, have also been explored for SEM images \cite{lazar2015sparsity, trampert2018should, chaudhary2019line, lee2014improving}. An extended list of all denoising approaches is presented in Table \ref{tab:denoising_methods}. Simple measures like SNR and Structural Similarity Index Measure (SSIM) or noise variance can be used to evaluate SEM image quality \cite{sim2011image, thong2001single, sim2013noise, kamel2004image}.

\begin{table}[t]
    \caption{Denoising algorithms used on SEM images. Values in Denoised Image are improvement (in \%) over original raw SEM images for that specific quality metric. Algorithms tested on IC SEM images are marked as either Yes(Y) or No (N) in the IC column. All references use different testing data.}
    \label{tab:denoising_methods}
    \centering
    \scriptsize
    \begin{tabular}{|l|c|c|c|c|c|}
    \hline
    \multicolumn{1}{|c|}{Algorithm} & \multicolumn{2}{|c|}{Raw Image} & \multicolumn{2}{|c|}{Denoised Image} & IC\\ \cline{2-5}
    & SNR & SSIM & SNR & SSIM & \\
    \hline
          Adaptive Weiner \cite{mazhari2016suppression} & - & 0.624 & - & 18.5 & N\\    
NLM \cite{mazhari2016suppression} & - & 0.624 & - & 7.2 & N\\
Hysterisis \cite{mazhari2016suppression} & - & 0.624 & - & 24.2 & N\\
Wavelet \cite{chaudhary2019line} & 9.96 & - & 58.6 & - & Y\\
BM3D \cite{lazar2015sparsity} & 10.00* & - & 95.4 & - & N\\
K-SVD \cite{lazar2015sparsity} & 10.00* & - & 108.4 & - & N\\ 
GOAL \cite{trampert2018should} & 25.57 & 0.73 & 6.5 & 22.3 & N\\
Super Resolution \cite{trampert2018should} & 25.57 & 0.73 & -0.01 & 12.12 & N\\
Image Inpainting \cite{trampert2018should} & 25.57 & 0.73 & 22.6 & 21.9 & N\\
DLNN (SEMNet) \cite{chaudhary2019line} & 9.96 & - & 164.6 & - & Y\\
         \hline
    \end{tabular}
\end{table}

\vspace{5pt}\noindent\textbf{Segmentation:} This step involves the separation of structures in the IC image based on some qualifying fact. In SEM images, this would be the material represented by grayscale pixel intensity. Segmentation algorithms can be supervised, unsupervised or interactive. Supervised segmentation approaches, such as the Convolutional Neural Network (CNN) based deep learning approaches require massive amounts of manually ground-truthed image data \cite{hong2018deep, cheng2018hybrid}. This massive undertaking requires considerable time and resources. It is hard to replicate with the dataset being private and may not generalize to other ICs. The unsupervised approaches are based on generalizable features that can be found in the same IC or across ICs. For instance, the technique developed by \cite{cheng2018hierarchical, cheng2018hybrid, cheng2019global} relies on the fact that polysilicon structures and metal layer traces can be generated by simple Manhattan geometry contours. Interactive approaches require the operator to guide the segmentation \cite{das2009semiautomatic}. Another technique relies on using frequency-based texture signatures for different materials to segment out IC structure across multiple layers \cite{wilson2020lasre}. Simple image processing techniques have also been explored for segmenting SEM images \cite{doudkin2005objects, wilson2020histogram, lee2008effective, lagunovsky1998recognition}. Segmentation accuracy is measured in pixel accuracy, F-measure, and Intersection-over-Union (IoU). However, there are no metrics that specifically look for accuracy in electrical connectivity. A list of segmentation approaches and their reported accuracy is summarized in Table \ref{tab:sem_segmentation}.

\begin{table}[t]
    \caption{Segmentation algorithms utilized for segmenting IC SEM images. Papers reporting multiple accuracy values is marked with a range operator ($\rightarrow$)}
    \label{tab:sem_segmentation}
    \centering
    \scriptsize
    \begin{tabular}{|l|l|}
    \hline
    \multicolumn{1}{|c|}{Algorithm} & Accuracy\\
     & (in \%) \\
    \hline
    Graph-Cuts \cite{das2009semiautomatic} & 91.00\\ 
    Morphology and Semantics \cite{doudkin2005objects} & 90.00\\
    Watershed \cite{lee2008effective} & 94$\rightarrow$98\\
    Hybrid K-Means and SVM \cite{cheng2018hybrid} & 95.83\\
    Hierarchical Multiclassifier System \cite{cheng2018hierarchical} & 91.45\\
    Fuzzy C-means \cite{cheng2018hybrid} & 26.72 \\
    K-means \cite{cheng2018hybrid} & 26.72\\
    Recursive Multilevel Thresholding \cite{cheng2018hybrid} & 35.25\\
    Global Template Projection and Matching \cite{cheng2019global} & 88.54\\
    LASRE \cite{wilson2020lasre} & 86$\rightarrow$99\\
    Deep Neural Network \cite{hong2018deep} & 98.00\\
    \hline
    \end{tabular}
\end{table}

\begin{figure}
    \centering
    \includegraphics[width=0.5\columnwidth]{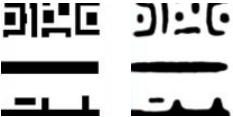}
    \caption{Edge distortion caused by limitations in the mask generation process in transition from the GDSII layout (left) to the silicon wafer (right) \cite{shao2020ic}}
    \label{fig:edge_effects}
\end{figure}

\vspace{5pt}\noindent\textbf{Vectorization: }Finally, the vectorization stage converts the image into a bunch of polygons. The idea behind this module is to recover design files as close to the original die layout as possible. This specific step enables the use of commercial off-the-shelf tools allowing smoother transition between the annotation and the netlist extraction modules. This step further serves in suppressing edge noise between materials (discussed in detail with EWR/LWR for FA \cite{pearce1987noise, constantoudis2013effects, midoh2005statistical}) and compressing the amount of data in the image. Vectorization has been achieved through simple edge following algorithms, supervised learning via XGBoost, and even deep learning \cite{blythe1993layout, quijada2018large}. It should be noted that edges in the original layout are converted to simple curves in the SEM images (see Figure \ref{fig:edge_effects}). This is due to the mask generation process done for etching the IC onto the wafer during manufacturing and not an artifact introduced by the imaging modality or any sample preparation step \cite{lippmann2020verification}. 

The annotation block is critical to several hardware assurance measures. Being in the middle of the RE workflow, the amount of errors accumulated at this stage is far lower than further down the line. Several approaches use segmented images of the IC layers for detecting Trojans. \cite{bao2014application, bao2015reverse} and \cite{nasr2017efficient} detect metal layer Trojans using classifiers trained on synthesized images. The former two works hand-craft shape features, while the latter one extract histogram of oriented gradients (HOG) features to represent discriminatory characteristics of gates. \cite{vashistha2018trojan} and \cite{shi2019golden} apply the Fourier shape descriptor to detect modifications on cells' appearance from SEM images of the IC's active layer. The classifier training in \cite{vashistha2018trojan} is conducted on a pre-collected IC SEM image dataset, and the training in \cite{shi2019golden} is on SEM images of on-chip electrically authenticated gates. 

Other approaches utilize SEM images of the golden IC or layout and simple correlation analysis or template matching to make sure that the sample is Trojan free \cite{vashistha2018detecting, courbon2015high}. Additionally, some approaches for avoiding Trojan insertion watermark the active layer with highly reflective filler cells. Their combined reflectance signature can be used to make sure that the IC is Trojan-free \cite{zhou2020hardware}. A similar approach uses thermal maps and deep learning to detect Trojans \cite{wen2020combining}. Being segmented, hardware assurance measures performed at this stage can also be free of the influence of localized variation in pixel intensity and noise, which has been demonstrated in the above works. It should also be noted that variants in Trojans, such as the parametric Trojans, can only be detected at this stage in RE. The only disadvantage at this stage is the overhead in storing and processing full-scale images.


\subsubsection{Netlist Extraction}

\begin{figure}
    \centering
    \includegraphics[width=0.9\columnwidth]{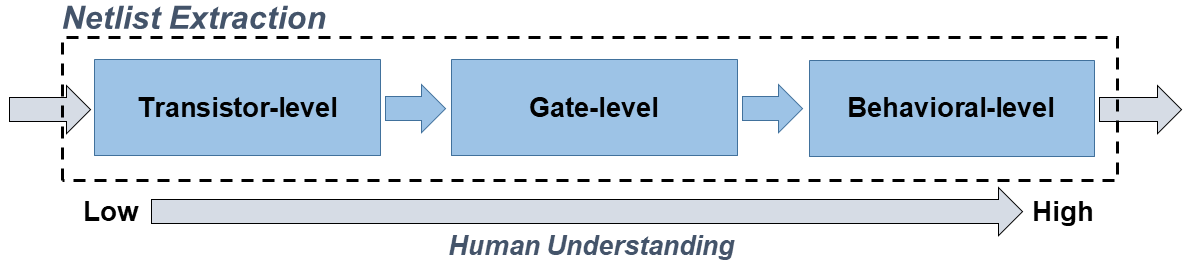}
    \caption{Expanded workflow for the netlist block. Operator understandability and intrepretability of the RE netlist increases from transistor-level towards behavioural-level netlist}
    \label{fig:netlist_flow}
\end{figure}

This stage involves the transition from the recovered layout into a netlist. As illustrated in Figure \ref{fig:netlist_flow}, this representation can be obtained at various levels of abstraction: transistor, gate and behavior-levels. This stage has the likelihood of having the most errors in the entire RE workflow. In contrast to popular belief, netlist extraction and debugging take more time and manual operator interaction than imaging and annotation modules. With present node technologies integrating several billion transistors in an IC die, it is reasonable to assume that there a billion opportunities to make mistakes. This is the underlying reason behind the existence of several hardware assurance approaches at the annotation module. However, in order to ensure that the hardware is completely Trojan free (like small scale Trojans with highly unique activation sequences) and to pave a path towards re-engineering obsolescent ICs, this stage is essential. In addition, there are several counter-RE approaches implemented at this stage, making error-free RE a daunting task. On a side note, it should be understood that RE is almost always performed on digital elements in the circuit. Large analog elements are extracted manually by the operator and decoded by an SME. 

\emph{Transistor-level} netlist can be generated for both digital and analog circuits. But it is predominantly done in analog circuits. The majority of the exemplary work in digital IC RE skips straight to the gate-level netlist. The transistor-level netlist is obtained either by simple image processing techniques like template matching and manual operation interaction. A transistor is detected at the overlap of diffusion
and polysilicon layers in the CMOS technology \cite{bourbakis2002knowledge}. The width for the n-type CMOS gate is assumed to be smaller than the p-type CMOS gate \cite{lippmann2020verification}.  

\emph{Gate-level} netlists operate by combining several transistors into logic gates. In most cases, the standard cell library that defines the gates are not available. Hence, with the help of Subject Matter Experts (SME), the operator has to manually identify each cell and use image-based correlation to generate a standard cell library \cite{6808796, kimura2020decomposition, nohl2008reverse}. Custom feature descriptors are also explored for improving cell template localization in large scale images \cite{dura2017fast}. This is a time-consuming process. Recently, an approach has been introduced that uses information from the contact layer to identify the standard cell library automatically \cite{wilson2019novel}. However, this approach assumes that no obfuscation has been performed on the cells. The amount of prior information at this stage significantly help speed up the RE process. For instance, if the standard cell library is known beforehand, the process of netlist recovery would be easier. For entities that have access to the golden netlist, a simple comparison can assure that the device under test is Trojan free. The validation of both the extracted transistor and gate-level netlist is done through simulations and design rules checks. Electronic Design Automation (EDA) tool suites, such as those provided by Cadence, are used to perform these checks \cite{quijada2018large}. A sample time frame for netlist extraction reported in a case study is shown in Table \ref{tab:time_netlist_extract}.

\begin{table}[t]
    \caption{Breakdown of logic gate placement and debugging netlist time frame (in minutes) for a 130nm node technology IC (Die Area: 558,000 $\mu$m\textsuperscript{2}) \cite{kimura2020decomposition}. Debug time takes a significant portion of the time frame for the netlist extraction process. Netlist verification and validation processing times vary from 30-206 minutes in these cases}
    \label{tab:time_netlist_extract}
    \scriptsize
    \centering
    \begin{tabular}{|c|c|c|c|}
    \hline
        Mode & Manual & Automatic & Comment\\
        \hline
        Placement & 168 & 0.8 & Design Data Available\\
        Debug & 106 & 0 & Design Data Available\\
        Placement & 59 & - & Design Data Not Available\\
        Debug & 39 & - & Design Data Not Available\\
    \hline
    \end{tabular}
\end{table}

A \emph{behavioral-level} netlist is generated to understand the functionality of the circuit itself. This facilitates obtaining insights on the design, especially for competitive intelligence and IP protection. Module functionality and boundaries are recovered at this step. Another one of the core reasons for RE, the recovery of schematics, is performed at this stage. For obsolescent ICs, the RTL description can be extracted from the gate-level netlist and synthesized into contemporary node technology, which can be manufactured in a fab with or without changes to the existing design. An exemplary work has been performed for this specific purpose \cite{kimura2020decomposition}. However, an in-house tool has been used for extracting RTL description from the gate-level netlist.
Bao et al. list three different entities interested in RE: circuit designers, foundries, and IP vendors \citep{bao2018reverse}. Another entity that may have an interest in RE is an external party. Design files, being privileged information, are not available to all entities. The spectrum of information goes on a monotonically decreasing scale from the circuit designers having access to the most information to the external party having at most access to the datasheet of the IC. 

As a reflection of this fact, there is a sea of methods for handling information at the behavioral netlist-level. In situations where the netlist needs to be verified, simple and scalable graph isomorphism based approaches identify sub-circuits in the flattened gate-level netlist to generate higher-level functional/behavioral modules \cite{cheremisinov2020subcircuit, li2012reverse, yang2003frosty, chisholm1999understanding, chang2001novel, li2013wordrev, terem2004pattern, rubanov2003subislands, subramanyan2013reverse}. Stochastic algorithms assume that structural similarity implies functional similarity and exploit this principle for extracting sub-circuit when the exact definition of the sub-circuit is not available \cite{baehr2020machine, baehr2019machine, zhang2003fuzzy, zhang2006speeding, shi2012extracting, fyrbiak2019constructive}. Probabilistic and fuzzy approaches are essential for error-tolerant matching/analysis in corrupted or obfuscated netlists. Machine learning approaches such as genetic algorithm and embedding space-based approaches have also been investigated \cite{vijaykrishnan1996subgen, cakir2018reverse}. There are some outlier detection methods that work on providing further insight into the functionality of the IC. Recovering state machine from the Input/Output pins by brute-forcing is one such approach \cite{smith2013non}. Recently, pioneering work has been done for the adoption of deep learning methods like graph convolution neural networks, for behavioral analysis on transistor-level netlists \cite{kunal2020gana}. 

\vspace{5pt}\textbf{Countermeasures against netlist extraction: }There are several obfuscation methods available for securing netlist-level information. Some techniques rely on layout-level camouflage, making the layout hard to decipher visually. Some approaches rely on gate-level camouflage, making functionally distinct logic gates look very similar under visual observation \cite{cocchi2014circuit, rajendran2013vlsi}. Since a lot of optimizations are employed in the stages between IC design and manufacturing, some approaches rely on overriding these optimization to generate functionally identical logic gates with very different physical/layout realizations \cite{gomez2018standard, gomez2019defeating}. It should be noted that deobfuscation approaches have also been proposed against these measures \cite{el2015integrated}. However, they do hinder the RE workflow in a significant way. For instance, with SME manually evaluating and extracting logic gates for generating standard cell libraries, a camouflaged gate may be accidentally missed or miss-classified. Arguments related to such human factor errors have been discussed in earlier works \cite{fyrbiak2017hardware}. 

Some other approaches for hardware assurance are also performed at this level. These include logic locking and split manufacturing \cite{gu2018cost}. Since they do not directly influence the RE process, these approaches are considered to be out of scope for this tutorial. Various studies that look into these approaches in detail, see, e.g., \cite{knechtel2019protect, tan2020benchmarking}. Trojan detection methods utilize the gate-level netlist to look for anomalies. As indicated in Figure \ref{fig:Taxonomy_trust}, Trojan detection can be performed with or without the golden data. Direct comparison of netlists, both transistor and gate-level, is resource-hungry. This motivated the development of Trojan net features. These features can be extracted from the golden netlist, and Trojan detection performed using several classifiers. Random forests, artificial neural networks, and SVMs have been explored in this context \cite{hasegawa2016hardware, hasegawa2017hardware, hasegawa2017trojan, chen2019hardware}. When direct access to golden netlists is not available, unsupervised cluster analyses detect Trojan infected circuits as outliers \cite{salmani2016cotd, cakir2015hardware}. Another approach using test patterns generated by Automated Test Pattern Generation (ATPG) tools has also been suggested \cite{sarkar2019automating}. Since non-invasive Trojan detection methods is out of scope of this tutorial, interested readers are directed to other detailed surveys on the topic \cite{elnaggar2018machine, liakos2019machine, li2016survey}. Table \ref{tab:trojandetection} lists hardware assurance approaches on both the annotation and netlist extraction modules.

\begin{table}[]
\caption{Performance of Trojan Detection Works}
\label{tab:trojandetection}
\scriptsize
\begin{threeparttable}
\begin{tabular}{|c|c|c|c|c|c|c|c|}
\hline
\multirow{2}{*}{Detection} & \multirow{2}{*}{Work}   & \multirow{2}{*}{Golden data}    & \multicolumn{5}{c|}{Performance\tnote{1}  }         \\
\cline{4-8}
 &   &  & Accuracy   & FPR   & FNR    & TPR    & TNR \\
\hline
\multirow{8}{*}{Image-based}    & \cite{bao2014application, bao2015reverse} & Golden layout   
                                & 99.9\%     &   -   &    -    &      -   &    -     \\
                                & \cite{nasr2017efficient}      & Golden layout  
                                & 89.0\%      & 7.7\%   &  -  &  - &   -  \\
                                & \cite{zhou2020hardware}                   & Golden layout   &    -   & 0.5\%    & 1.2\% & -  & -   \\
                                & \cite{wen2020combining}                   & Golden layout   & 98.2\%     &   -   &  -  &  -   &  -  \\
                                
                                & \cite{shi2019golden}     & Golden layout  
                                & 100\%    &  -   & -  & -  &  -   \\
                                & \cite{vashistha2018trojan}    & Golden IC   
                                & 98.0\%    & 2.0\%   &  -  & -  & -  \\
                                & \cite{courbon2015high, vashistha2018detecting}     & Golden IC 
                                & 100\% visually   &  -  & -  & -   &  -  \\
\hline
\multirow{6}{*}{Netlist-based}  & \cite{hasegawa2016hardware}   & Golden netlist
                                &   -   &  -   &  -  & 96.8\% & 66.3\% \\
                                & \cite{hasegawa2017hardware}  & Golden netlist  
                                &    -   &  -  & -  & 85.0\%    & 70.0\%    \\
                                & \cite{hasegawa2017trojan}  & Golden netlist 
                                & -  &  -   &  -   & 70.3\% & 99.7\% \\
                                & \cite{chen2019hardware}  & Golden netlist  
                                & 99.8\%     &  -   & -   & 89.8\% & 99.9\% \\
                                & \cite{cakir2015hardware}   & Not required 
                                & 97.5\% & -  & 70.3\% & - & - \\
                                & \cite{salmani2016cotd}     & Not required   
                                & -  & 0\%     & 0\%    &   - &  -  \\
                    
\hline
\end{tabular}
\begin{tablenotes}
\item[1] Table presents the averaged reported performance or the best averaged performance if multiple averaged performance are available. FPR: False Positive Rate, FNR: False Negative Rate, TPR: True Positive Rate, TNR: True Negative Rate.
\end{tablenotes}
\end{threeparttable}
\end{table}

\vspace{5pt}\noindent\textbf{Available tools: }Unlike the imaging and annotation modules, the netlist extraction module is well-studied in the literature. There are several benchmark datasets available such as ISCAS \cite{hansen1999unveiling}, TrustHub \cite{salmani2013design}, Synopsis Academic Datasets \cite{goldman201332, goldman2009synopsys}, OpenCores \cite{opencores2018}, IWLS \cite{iwls2005}, OSU35 \cite{osu35} and FreePDK \cite{free45nm} that help generate sample data for hardware assurance applications such as Trojan detection.

\begin{table}[t]
    \caption{Commercial and open-source tools available for RE related tasks}
    \label{tab:tools}
    \centering
    \scriptsize
    \begin{tabular}{|l|l|}
    \hline
         Name & Purpose \\
        \hline
         ARES \cite{thomasimpact} & Segmented image to GDSII layout \\
        PIX2net\footnote{http://micronetsol.net/} & Annotation \\
        Degate \cite{vijayakumar2016physical} & Gate recognition and annotation \\
        GDS-X \cite{quijada2018large} & Segmentation-to-polygon conversion \\
        CircuitFinder \cite{kim2018fast} & Annotation \\
        ICWorks \cite{wang2017probing} & Gate netlist extraction \\ 
        HAL \cite{wallat2019highway} & Gate netlist manipulation \\
         \hline
    \end{tabular}
\end{table} 

In recent years, several tools have been introduced that can significantly reduce the workload associated with RE. A list of these tools can be found in Table \ref{tab:tools}. One of the three main reasons for performing RE is achieved at this level of abstraction: detection of anomalies in the design for hardware assurance purposes. Hardware assurance approaches for post-CMOS technologies are out of scope of this tutorial.  Interested readers are directed to recent work by Knechtel for more details \cite{knechtel2020hardware}. 

\subsection{Problems Associated with Applying ML for PCB RE}
Here, similar to our discussion on how ML can be useful for IC RE, annotation, and netlist extraction techniques carried out in the context of PCB RE are explained. 
\subsubsection{Annotation}

Machine learning combined with image analysis has proven invaluable for quality control and hardware assurance in the PCB manufacturing industry, enabling automated defect detection and visual inspection to a certain degree \cite{chavan2016quality, chaudhary2017automatic, kaur2014detection,huang2015detection, tian2014application, leta2008computer,  benedek2012solder, scaman1995computer,wang2016machine, kim2012automatic, gaidhane2018efficient, zhu2018printed, rehman2019automated, moganti1998segmentation, hassanin2019real, chavan2016quality, huang2015automated, kuo2019data, zakaria2020automated}. 
Several studies have applied image subtraction to compare a golden, reference image of a PCB design or schematic to PCBs under quality check \cite{leta2008computer, chavan2016quality, chaudhary2017automatic, kaur2014detection, tian2014application}. Other more complex approaches for defect detection in PCBs involve modeling-based methods, such as evaluating the roundness of drilled vias \cite{wang2016machine} or using multi-marked point processes for solder paste defect detection \cite{benedek2012solder}.

The first phase of the annotation for PCB deals with identifying whether the PCB is imaged optically or via X-Ray CT and whether external RE or internal RE should be considered, respectively. Clearly, the
denoising required for each of these modalities is different as the noise sources vary from one to another. For external RE, this step includes generating the design's BoM by extracting the components, vias, traces, silkscreen annotations, etc. Internal PCB RE mainly concerns extracting the internal routings and connections of traces and vias, with the main difference being the noise involved. The final step for both external and internal RE infers the purpose and functionality of the circuit, sub-circuit, or system.

\begin{figure}[t]
    \centering
    \includegraphics[width=0.5\columnwidth]{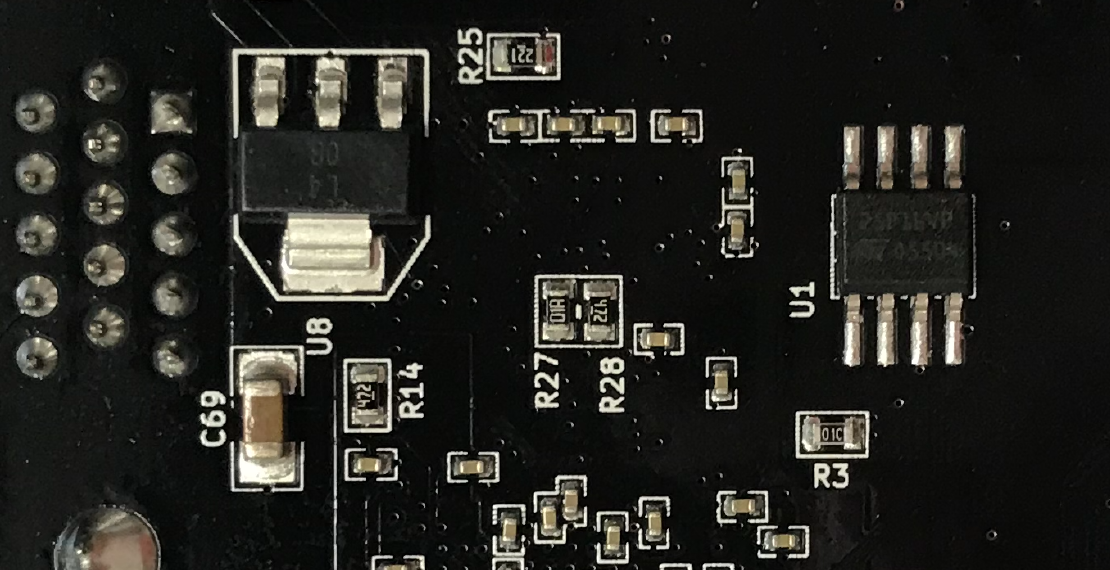}
    \caption{An example of extracting the same color PCB component from the PCB board.}
    \label{fig:extColor}
\end{figure}

\begin{figure}[t]
    \centering
    \includegraphics[width=0.5\columnwidth]{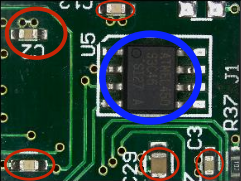}
    \caption{Optically imaged PCB segment highlighting components: Capacitors (Red) and IC (Blue)}
    \label{fig:extComp}
\end{figure}

\vspace{5pt}\noindent\textbf{External RE: }When applying external RE, the desired features to be annotated are the components, text, logo markings, vias, traces, etc. 
From an image analysis perspective, the extraction of such features falls within the scope of object detection, classification, and identification. 
Existing algorithms for achieving these goals are well-developed for natural scene images \cite{viola2004robust, redmon2016you}.  
Nevertheless, adapting these algorithms for external PCB RE-related applications remains challenging.   
The components on the surface of a PCB vary in size and color depending on their functionality and packaging. This could challenge the annotation process. First, the lighting conditions and color of PCB surfaces may impact the extraction performance. Traditional image analysis methods convert an RGB image to another color map to address the lighting variance \cite{lecun2004learning, sural2002segmentation}; however, this cannot prevent shadowing on PCB surface due to the existence of tall components, which may result in an error of using color-based segmentation methods.  
Moreover, the complications when the color of a PCB's surface is similar to the color of components has not been addressed completely (see Figure \ref{fig:extColor} where the colors of the surface of the PCB and components are all black). Besides, in an optically imaged PCB, text markings, traces, and vias are packed tightly compared to the objects in a natural scene image. This increased image complexity challenges feature localization,
especially in a densely populated design. In addition, with the advances in technology, the size and shape of components become smaller, and their placements/orientations in the design are decided by the designer. Therefore, neither specific rules nor the encoding that hold from one design to another across even one generation can be defined. 

While the above challenges should be addressed, there is a plethora of side information on the board itself that can be leveraged to make the 
annotation task easier. The on-board text marking near a component represents the type of components (see Figure \ref{fig:extComp} where C2/C3/C29 are capacitors and U5 is an IC). Those markings can be used as ground truth, which provide machine learning classifiers with either the labels or additional features during classification.
Although there are substantial applications for these markings, extracting them is particularly challenging. The Optical Character Recognition (OCR) is the most widely used tool for text recognition \cite{smith2007overview}, but its performance is not stable in the case of PCB inspection. The markings are etched or printed on PCB boards using a variety of materials and colors, and they vary in orientation, which degrades the performance of the OCR engine \cite{li2014text,gang2020coresets}. This stresses the importance of having a robust text recognition system for external PCB RE applications. ~\cite{gang2020coresets} looked further into these data collection issues by investigating which features of components or board text-markings can form a coreset (small weighted summaries of a larger set of data) to more efficiently and effectively train deep learning text-extraction models. Using data gathered from real-world PCB production sites, they developed a series of coresets with varying combinations of illumination variance or shape orientation and evaluated their effectiveness on training the standard ResNet~\cite{he2016identity} deep neural network model. With a top reported accuracy of 94.67\% from experiments, they then determined that the most important features for accurate text-extraction were the shape and orientation of the characters.   

In addition to the above-mentioned markings, information can be derived from the traces and vias on the board.
The extraction and localization of traces and vias are critical because they determine the functionality and the performance of the board, which allow the validation of the system's integrity. Existing research on the detection of traces and vias has mainly focused on finding defects and usually uses the bare board to illustrate the problem \cite{dave2016pcb, tang2019online}; however, the proposed approaches are not robust in practice when traces and vias are overlapping with components. 

\begin{figure}[t]
    \centering
    \includegraphics[width=0.5\columnwidth]{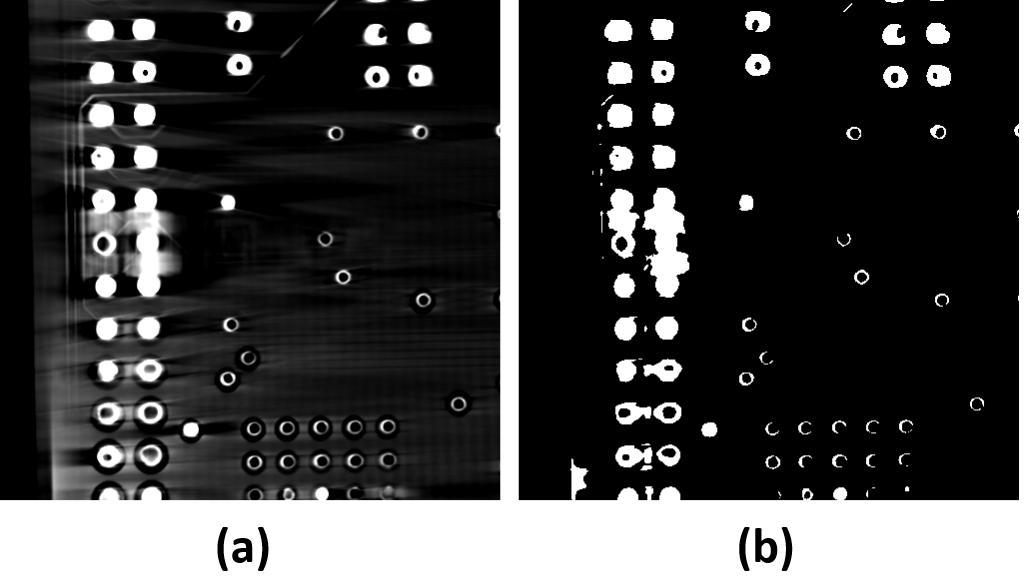}
    \caption{X-ray CT PCB: (a) Raw Image (b) Post Segmentation}
    \label{fig:xrayThresh}
\end{figure}

\begin{figure}[t]
    \centering
    \includegraphics[width=0.5\columnwidth]{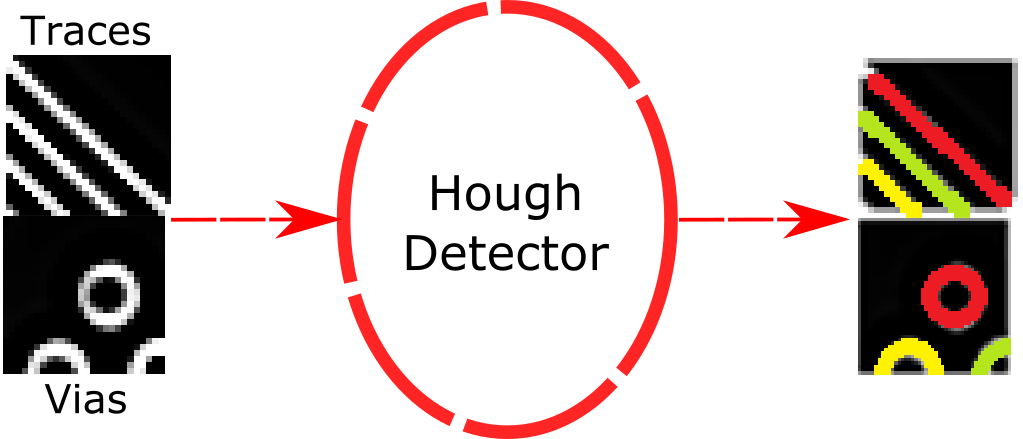}
    \caption{X-Ray CT trace and VIA detection using Hough line/circle detectors.}
    \label{fig:houghEx}
\end{figure}

\vspace{5pt}\noindent\textbf{Internal RE: }For internal PCB RE, the important design information is the vias and traces on a board and the layers throughout the board. The vias establish connectivity between the layers and are consistent from one layer to another, except in a rare case of a blind/buried vias in a space-constrained design. These blind vias are unique only to their respective layers as opposed to being at the same location throughout \cite{mcloughlin2008secure}. Traces provide the main discriminatory information for each layer and determine the connectivity of the vias throughout the board. 
As stated previously, the noise affects the X-Ray image quality, which particularly impacts 
feature extraction 
for annotation, when the traces and vias 
in internal PCB RE can be significantly altered due to the blur and high-z material noise artifacts. For example, vias in the raw image can be linked (see Figure \ref{fig:xrayThresh}~(a)) or their shape can be noticeably distorted (see Figure \ref{fig:xrayThresh}~(b)). 
The distortion may degrade the performance of the feature localization and identification.
Vias and traces within a PCB are mainly circles and lines (see Figure \ref{fig:houghEx}). The predominant class of algorithms for detecting these geometries, line/circle detection using model-fitting methods, is quite sensitive to the parameters and pixel intensities. Thus, it likely would require manual tuning from one sample to another to minimize the number of missing or falsely detected objects. This is neither practical nor ideal for automated PCB RE, where it should be noise-tolerant, generalizable, and scalable for multiple technologies and designs. 

\begin{figure}
    \centering
    \includegraphics[width=0.5\columnwidth]{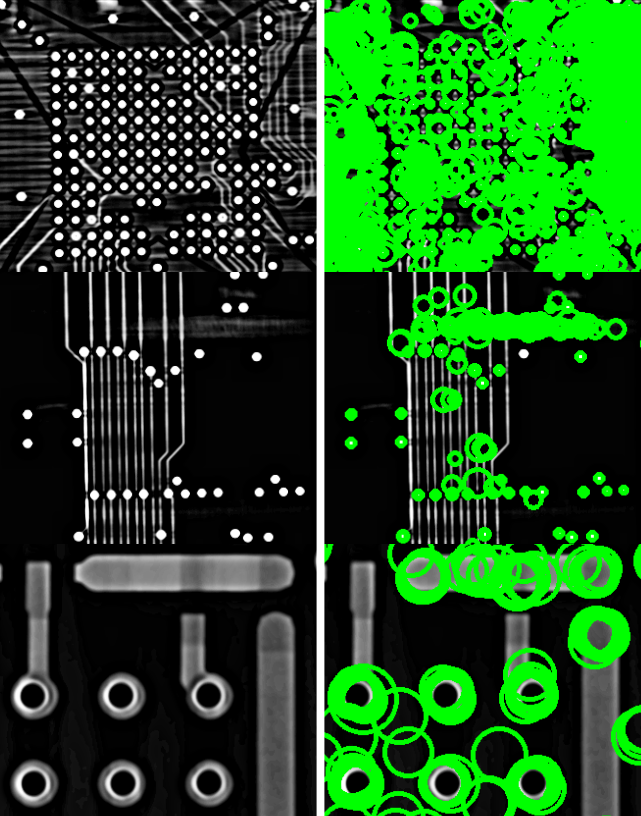}
    \caption{Hough circle detection false positives for x-ray CT imaged board samples}
    \label{fig:houghFP}
\end{figure}

The topic of automated via detection has been explored recently by~\cite{liu2018automatic, botero2020Vias} with both leveraging the popular Hough circle transform~\cite{Kalviainen1995ProbabilisticComparisons} for detecting vias' circular shapes in X-ray CT images.~\cite{botero2020Vias} in particular combined Hough circle detection with an unsupervised iterative learning approach and PCB design domain knowledge to address the challenges seen from noise and design variance. 
Due to the known shortcomings of Hough circle detection, namely sensitivty to noise and propensity for false positives(See Figure~\ref{fig:houghFP}), they utilize a series of stages to iteratively screen out false positives and learn the properties of true positive vias unique to each board at the same time. To further emphasize the need for robust approaches that can learn online in an unsupervised fashion, the authors compare their results to the current state of the art for deep learning-based object detection, Mask R-CNN~\cite{he2017mask}. 

\begin{figure}
    \centering
    \includegraphics[width=0.5\columnwidth]{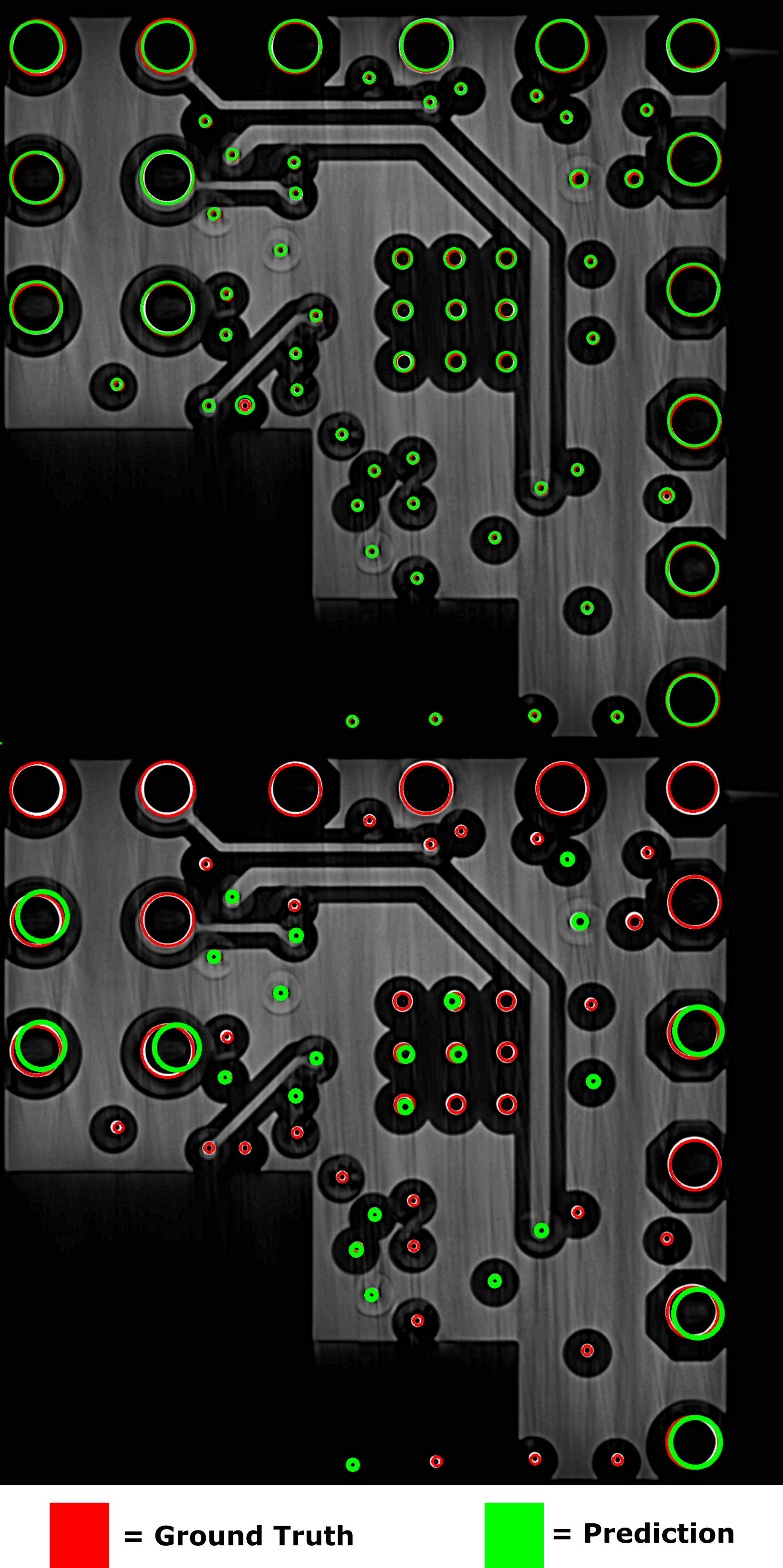}
    \caption{Automated Via Detection Results Test Board: ~\cite{botero2020Vias} Proposed Methodology (Top) Vs. Mask R-CNN (Bottom)}
    \label{fig:viaRes}
\end{figure}

Mask R-CNN(M.R-CNN) detects objects in an image while simultaneously generating a high-quality localized segmentation mask for each instance. The implementation of Mask R-CNN used in ~\cite{botero2020Vias} has been pre-trained on the MS COCO Dataset~\cite{lin2014microsoft}, and the pre-trained weights have been used to train a variation of the network to locate and identify vias. 
Table~\ref{table:via} clearly highlights how the authors' proposed automated via detection framework vastly outperformed training and implementing Mask R-CNN for the same tasks based on commonly-used image segmentation metrics. The authors attribute this vast difference in performance to their proposed framework's ability to adapt online to novel data as opposed to being reliant upon trained data, a known hindrance to deep learning methods. 
Figure~\ref{fig:viaRes} presents the results of the automated via detection with predicted vias in green and the ground truth highlighted in red in the same image.   


\begin{table}
\centering
\caption{Automated Via Detection Segmentation Mask Evaluation Results: ~\cite{botero2020Vias} Proposed Methodology Vs. Mask R-CNN (T.B. = Test Board, S3 = Spartan 3)\vspace{-5pt}}
\label{table:via}
\scriptsize
\begin{tabular}{|c|c|c|c|c|c|c|} 
\hline
\multicolumn{7}{|c|}{Automated Via Detection Segmentation Mask Evaluation}                                              \\ 
\hline
Board       & \multicolumn{2}{c|}{Average IoU} & \multicolumn{2}{c|}{Average DICE} & \multicolumn{2}{c|}{Average SSIM}  \\ 
\hline
            & ~\cite{botero2020Vias} & M.R-CNN      & ~\cite{botero2020Vias} & M.R-CNN       & ~\cite{botero2020Vias} & M.R-CNN        \\ 
\hline
T.B. & .877            & .564           & .930            & .617            & .973            & .638             \\ 
\hline
S3   & .730           & .511           & .820           & .567            & .949           & .847             \\
\hline
\end{tabular}
\end{table}

Trace detection in X-ray CT images has also drawn recent research interest~\cite{qiao2018wire}. Qiao et al. used a combination of deep convolutional neural networks (DCNN) with graph-cut models to semantically segment trace wires in a variety of X-ray CT imaging environments. 
The authors adopt a transfer learning approach to train their network due to traditional DCNN training requiring exorbitant and time-consuming manual pixel-level labeling of ground truth. 
They first trained their model for classification between the categories of trace, via, pad, and background 
followed by shifting the focus of the trained network from classification to semantic pixel-level segmentation after a series of architectural changes further outlined in~\cite{qiao2018wire}. They then use the DCNN segmented result as a probability prior for their graph cuts model that further refines the result in combination with low-level feature information from the original image. The authors compare their approach against two traditional segmentation methods (level set~\cite{yuan2010study} and superpixel~\cite{kai2016automatic}) as well as variants of their trained network with and without graph cut refinement. 

Evaluation of the method is performed based on segmentation metrics pixel accuracy(PA) and IoU, as well as classification metrics recall, precision, and F1-Score(See Table~\ref{table:traceWire}). The level set and superpixel-based methods both performed the worst because these methods perform segmentation based on color information or local texture and do not have the same powerful recognition ability as DCNN-based methods. However, the results of the trained DCNN methods are still coarse, with errors leaving room for improvement. These results both emphasize the need for more sophisticated segmentation methods and less training data reliant approaches due to the impracticality of requiring such extensive manual labeling, as mentioned earlier. 

\begin{table}
\centering
\caption{Trace wire segmentation performance comparison~\cite{qiao2018wire}\vspace{-5pt}}
\label{table:traceWire}
\scriptsize
\begin{tabular}{|c|c|c|c|c|c|} 
\hline
\multicolumn{6}{|c|}{Trace Detection Performance Comparison}  \\ 
\hline
Method     & PA    & IoU   & Recall & Precision & F1-Score    \\ 
\hline
Levelset~\cite{yuan2010study}   & .8246 & .5743 & .4585  & .3123     & .3715       \\ 
\hline
Superpixel~\cite{kai2016automatic} & .8074 & .6534 & .4792  & .6064     & .5353       \\ 
\hline
DCNN-GC~\cite{qiao2018wire}    & .9002 & .8463 & .8629  & .8655     & .8642       \\
\hline
\end{tabular}
\end{table}

\begin{figure}[t]
    \centering
    \includegraphics[width=0.5\columnwidth]{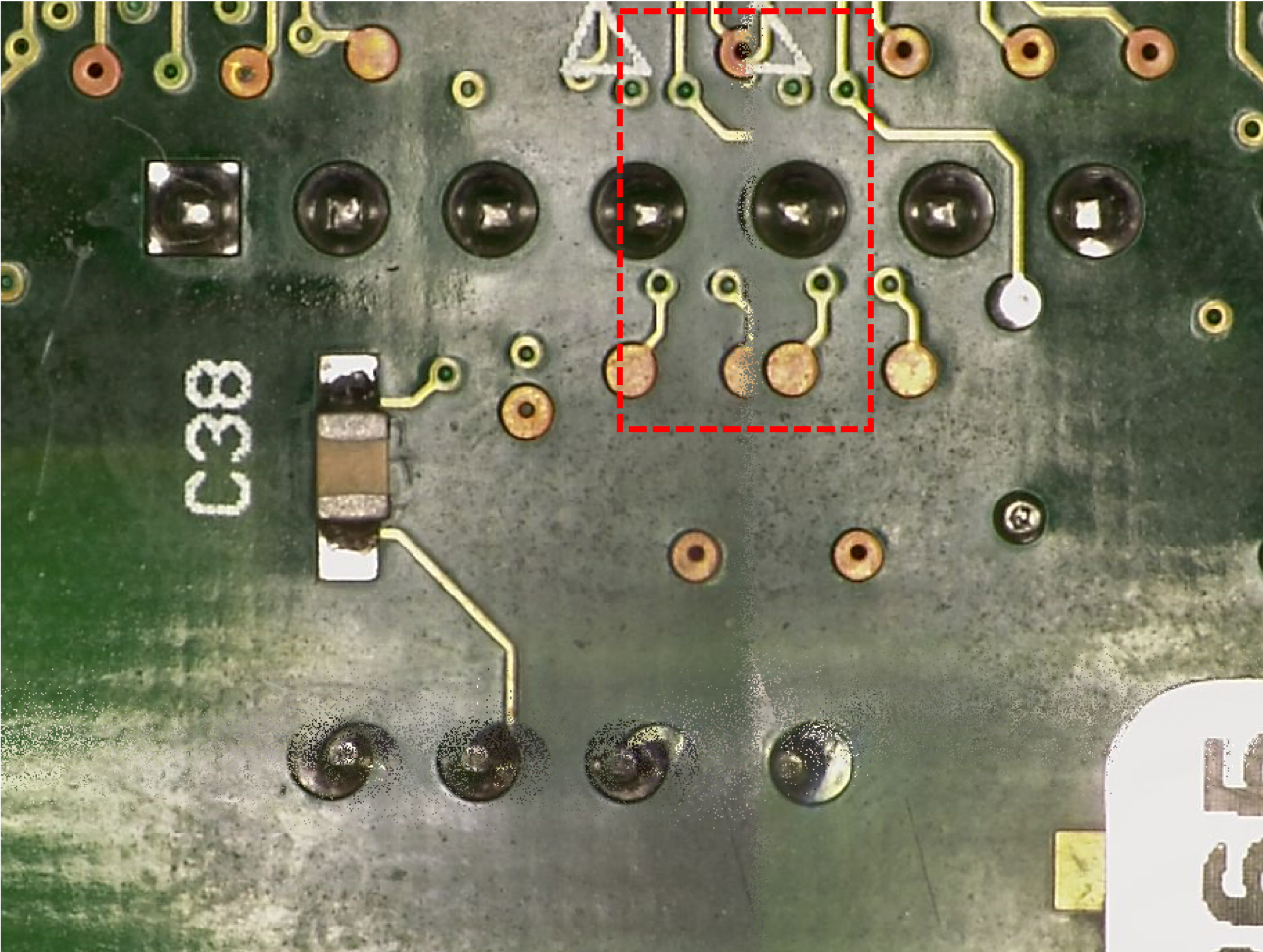}
    \caption{Misaligned external PCB image}
    \label{fig:unalign_external}
\end{figure}

\begin{figure}[t]
    \centering
    \includegraphics[width=0.5\columnwidth]{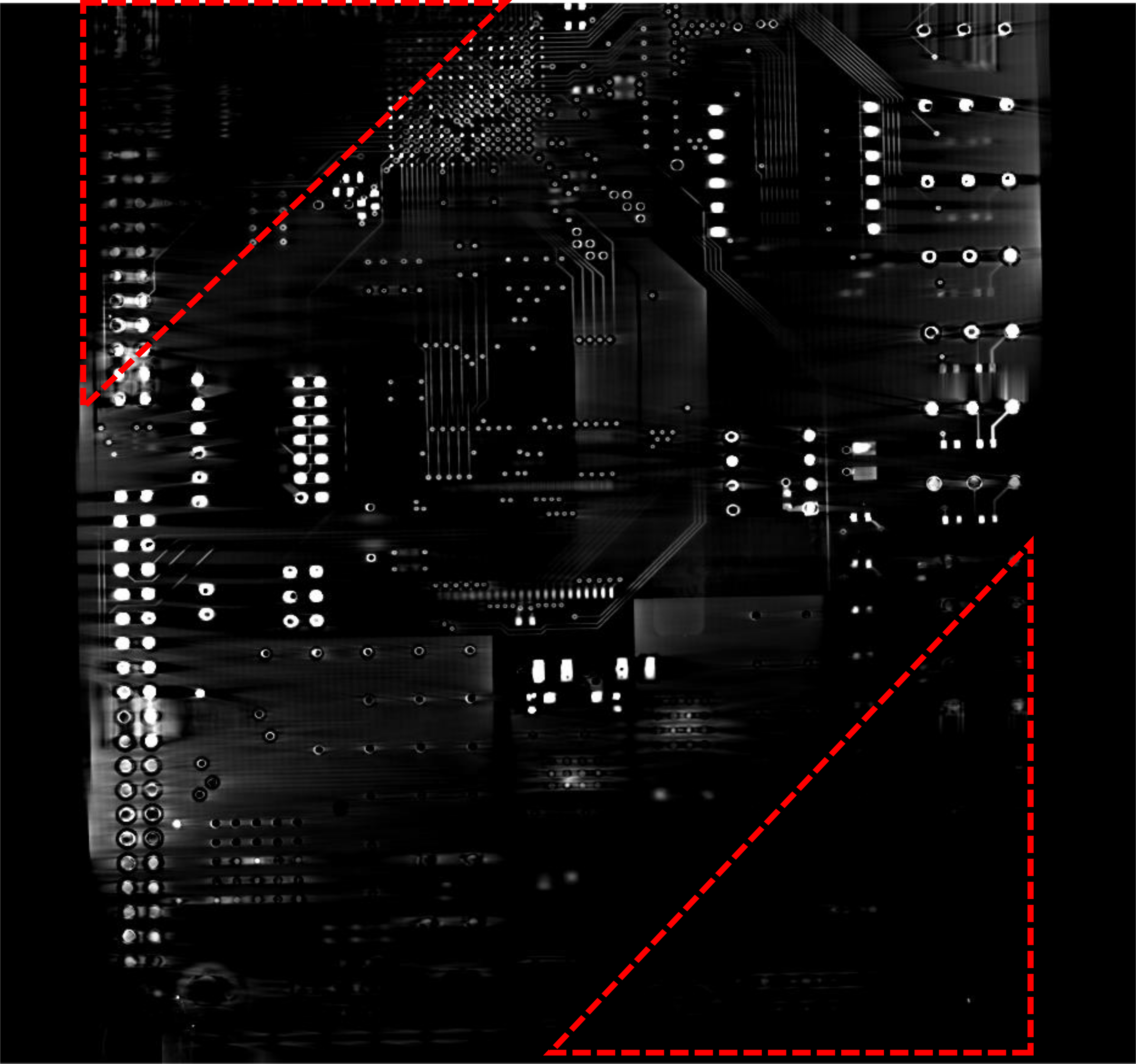}
    \caption{Misaligned Xilinx Spartan board X-Ray CT slice}
    \label{fig:unalign}
\end{figure}

In addition to the challenges mentioned above, improper alignment causes another difficulty that affects the feature extraction in both external and internal PCB RE cases. Due to the trade-off between details included in the features and the size of the features in external PCB imaging, some components cannot be captured completely in a single image, and thus we require aligned (stitched) images to extract the features. Although this matter has been dealt with by using the Charge-Coupled Device (CCD) camera and SEM imaging \cite{brown2007automatic, ma2007use}, the lighting condition and the number of assembled components can still lead to the stitching errors (see Figure \ref{fig:unalign_external}) in external PCB images. The same issue happens in the internal PCB RE. The aliasing effect in the slices that make up the 3D board sample (see Figure \ref{fig:xrayNoise}) is a byproduct of misalignment. This leads eventually to having the layers containing various amounts of feature information that may associate with a single layer or an adjacent one. Additionally, misalignment can also lead to another issue -- missing information (see Figure \ref{fig:unalign} where red dashed regions show areas of missing information). Therefore, it is important to not only be able to detect these features at each slice during feature extraction, but also address misalignment beforehand in order to localize and correspond the features to their correct layer. 

\begin{figure}
    \centering
    \includegraphics[width=0.75\columnwidth
    ]{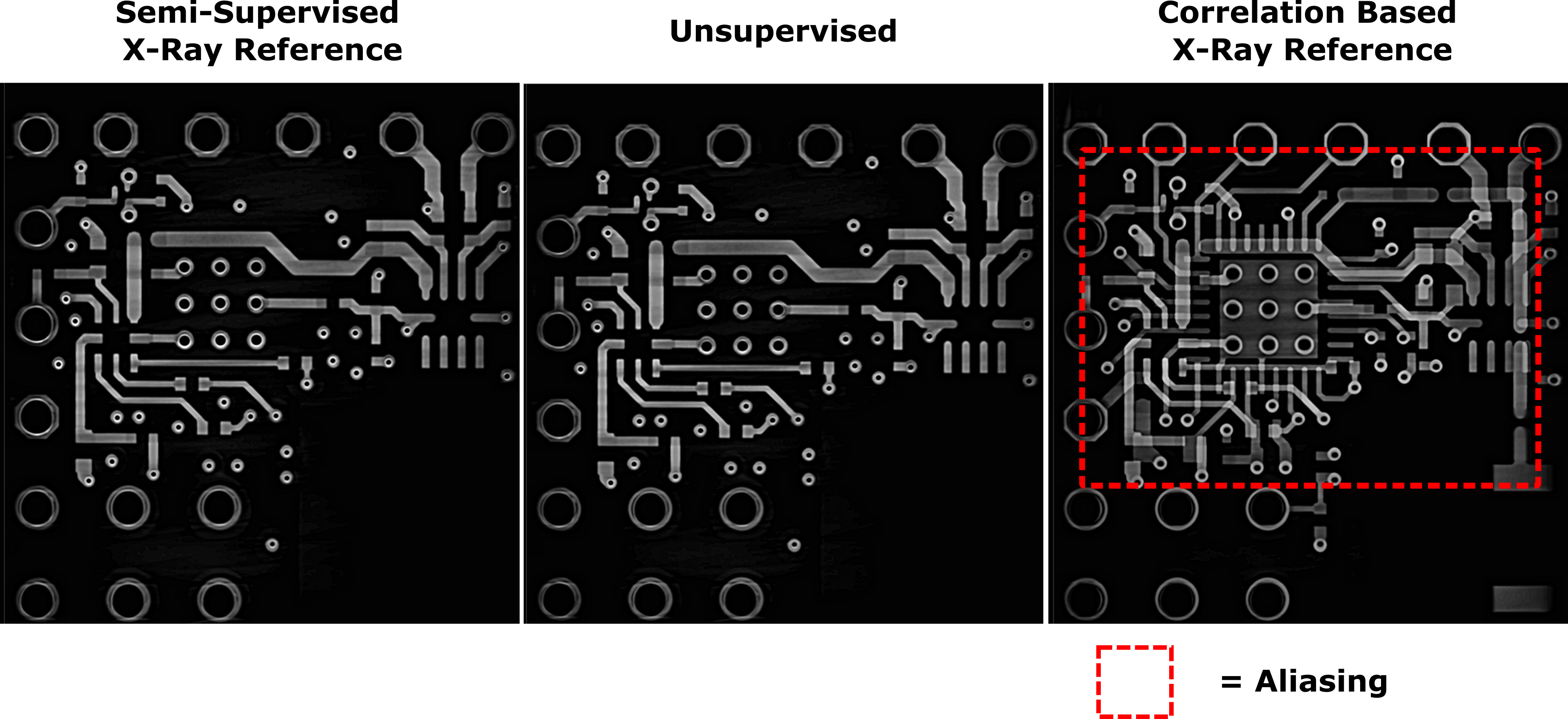}
    \caption{Layer Identification Results ~\cite{botero2020SemiSupervisedLayers} }
    \label{fig:semiSupeRes}
\end{figure}

There has been early work to address the misalignment/aliasing challenge~\citep{Botero2020APCBs} in internal PCB RE, as well as identifying what slices belong to what layers of a volumetric PCB stack depending on varying levels of a priori information~\cite{Botero2020APCBs, botero2020SemiSupervisedLayers}. First,~\cite{Botero2020APCBs} provided an algorithm that enables fully automated unsupervised 3D-stack alignment via iterative robust regression and registration based on the image gradient direction of the volumetric PCB stack. With these slices now aligned and registered, the data better facilitates slice-to-layer identification, which is key for internal PCB RE and determining which features correspond to which layers. ~\cite{Botero2020APCBs} then leverages the similarity of aligned adjacent slices in an X-ray CT imaged volumetric stack by using K-means clustering, in combination with exponentially windowed cross correlation, to identify what slices belong to what layers in an automated unsupervised manner, where the reverse engineer in practice has no reference design information. 

In ~\cite{botero2020SemiSupervisedLayers} the authors then explore the effectiveness of automated slice-to-layer identification under the assumption that the RE has reference design information available either in the form of CAD software provided digital images of the layers present in the board, or their X-ray CT imaged equivalent. However, the challenge for this task is that the provided reference design knowledge may not always be of the same imaging modality as the X-ray CT data. To address this heterogeneous image comparison challenge, the authors employed a feature based approach utilizing the spatial pyramid bag of visual words model~\cite{Lazebnik2006} in an iterative manner with various weighting schemes and then compare their results not only against the unsupervised methodology~\cite{Botero2020APCBs} but also against naive pixel intensity based approaches. 
Table~\ref{table:semiSupervised} summarizes the results across the different methods from a classification perspective as well as image quality assessment due to the desired result for subsequent RE stages being a fused singular image of the slices composing a layer. 

In particular, in ~\cite{Botero2020APCBs}, the importance of having domain knowledge of the X-Ray CT process for solving the slice-to-layer correspondence issue is highlighted. Specifically, slice location information is critical for unsupervised slice-to-layer identification due to the similarity of adjacent slices in an X-ray CT imaged volumetric stack. Unsupervised approaches tested in ~\cite{Botero2020APCBs} that did not utilize exponential windowing to emphasize location did not report a single instance of an F1-Score above 10\% for classification accuracy. Furthermore, ~\cite{botero2020SemiSupervisedLayers} improved the peak unsupervised results through its iterative feature-focused approach as opposed to na\"ive intensity-based approaches that performed particularly poorly in comparison (See Figure~\ref{fig:semiSupeRes}).

\begin{table*}
\centering
\caption{Average semi-supervised classification performance and image quality assessment compared across approaches~\cite{Botero2020APCBs,botero2020SemiSupervisedLayers}}
\label{table:semiSupervised}
\scriptsize
\begin{tabular}{|c|c|c|c|c|c|} 
\hline
\multicolumn{6}{|c|}{Average Classification Performance \& Image Quality Assessment}                         \\ 
\hline
Method                              & Precision & Recall  & F1-Score & Average Correlation & Average SSIM  \\ 
\hline
Semi-Supervised - Digital reference & 97.2\%    & 97.2\%  & 97.2\%   & .999                & .997          \\ 
\hline
Semi-Supervised - X-Ray Reference   & 98.34\%   & 98.34\% & 98.34\%  & .999                & .998          \\ 
\hline
Unsupervised                        & 90.51\%   & 90.51\% & 90.51\%  & .997                & .974          \\ 
\hline
SSIM Based Digital Reference        & 36.31\%   & 36.31\% & 36.31\%  & .868                & .816          \\ 
\hline
Correlation Based X-Ray Reference   & 70.70\%   & 70.70\% & 70.70\%  & .957                & .844          \\
\hline
\end{tabular}
\end{table*}

Deep learning for PCB RE is equally challenging. Although for external PCB RE, it may seem apt to apply deep learning for object detection/classification, it can easily fail. This is due to the fact that deep learning models are data-driven. More precisely, they can be confused when encountering unseen data from custom ICs developed in-house or by a third-party foundry, as opposed to those seen before (e.g., commercial, available ICs). Besides, the passive components have a variety of footprints that often overlap, and consequently, a correct classification cannot be achieved in a straightforward manner.~\cite{kuo2019data} explored the component detection/extraction topic with a deep learning (region proposal networks) and graph theory-driven approach. They reported a mean Average Precision (mAP) of 0.653 for their top model, leaving much room for improvement. They also highlighted that the main difficulties are the ambiguity/subjectivity of labeling data, necessary for training deep learning models. 

Furthermore, deep learning has been adopted in studies on internal PCB RE as well. Qiao et al. has suggested using a Deep Convolutional Neural Network (DCNN) with the graph cuts to achieve segmentation in PCB CT images \cite{qiao2018wire}. Although the performance is improved, when compared to the performance on natural scene images, the network still presents huge improvement potential. 
To apply deep learning in PCB RE cases, the authors of \cite{qiao2018wire} leveraged transfer learning in combination with image patches from the complete image to address overfitting and other challenges caused due to the limited size of the training database. 

Nonetheless, it is not clear how well this technique can generalize to the vast variety of board samples and whether training is required for every new sample. Even if this issue can be resolved by adding other board samples, this intensifies the labeling problem. Lastly, Botero et al. highlighted the challenges to quickly adapt existing deep learning solutions to these problems when utilizing the Mask R-CNN object detection network for via detection~\cite{botero2020Vias}. When the authors have utilized a pre-trained network, whose weights were further trained for vias, the performance is not only sub-par for the training dataset but has poor translation to a novel dataset not seen during training.

\subsubsection{Netlist Extraction}

\begin{figure}
    \centering
    \includegraphics[width=0.6\columnwidth]{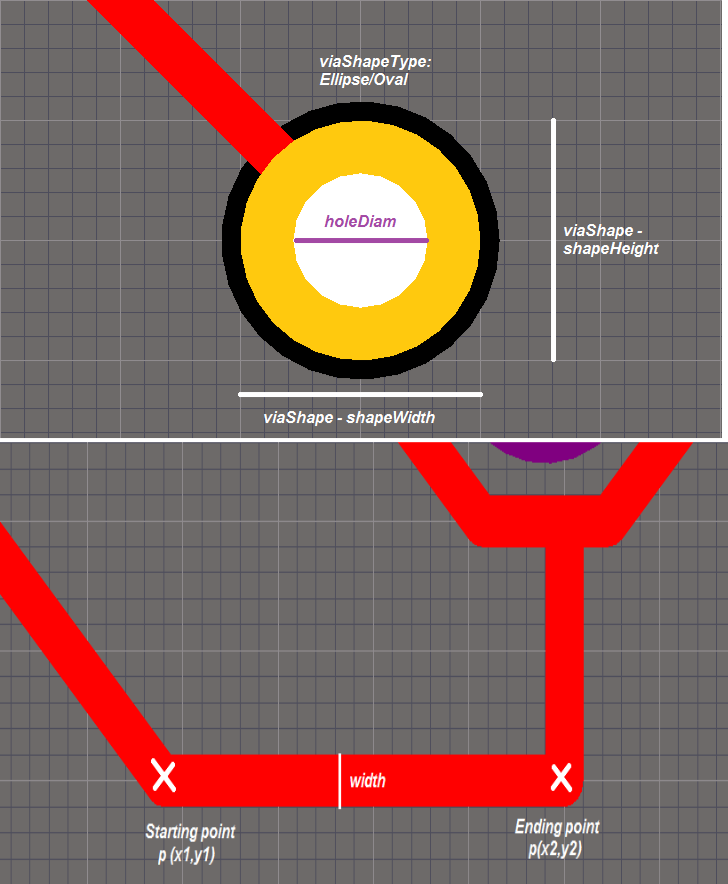}
    \caption{Via and trace vectorized features}
    \label{fig:vectViaTrace}
\end{figure}

The extracted and localized PCB features from the annotation stage are analyzed to generate a netlist that can be further manufactured. This requires the translation of the image features from the pixel domain to the geometrical domain (see Figure \ref{fig:vectViaTrace}). Once the BoM has been obtained from the PCB board, a software tool is applied to interpret the information by comparing that to a standard BoM for inspection, see, e.g., \cite{fu2007system, jin2000method}. Moreover, the systems, such as the work proposed in \cite{naveen1993automatic}, can generate a schematic from a netlist, enabling applications such as schematic verification, anomaly detection, replacement/upgrades, etc. However, these applications assume that the extracted features from PCB are correct, which may lead to inspection failure if the error is accumulated from earlier steps. Accordingly, it is crucial to do in-depth research in PCB feature extraction and analysis to obtain a more robust PCB RE scheme. 

Anti-PCB RE approaches are focused on making RE at the PCB-level prohibitively more expensive and time-consuming than it is worth. This is done in a variety of ways, with varying complexities. The work presented in \cite{mcloughlin2008secure} focuses on the design and implementation of a few passive components, more unmarked ICs, using custom silicon, and/or having cases of missing silkscreens. These methods have low to moderate overhead with regards to design cost and manufacturing impact. However, they only increase RE costs by a small margin. Instead, by utilizing a multi-layer board, blind and buried vias, and/or routing signals for the inner layers, the RE costs can be significantly raised \cite{quadir2016survey}. Additionally, obfuscation can also be utilized for anti-RE of PCBs \cite{guo2015investigation}, similar to its application for anti-RE at the IC level. Components referred to as permutation blocks can be used to hide the interconnects among the circuit components on PCBs.   

Existing research on how to counter X-Ray-enabled RE focuses on inserting X-Ray detecting sensors and using specific materials.  
X-Ray detecting sensors are devices embedded in a PCB design, being sensitive to X-Rays and react once a predetermined length of exposure to X-Rays has occurred. Afterward, the sensors signal a destructive measure to take place on the board to delay the RE process or simply act as an indicator that RE has taken place. Furthermore, high-density materials with high X-Ray attenuation factors, such as Zirconia powder, could be used throughout a board to reduce the quality of the X-Ray images. Using this material in a specific pattern throughout several layers can drastically affect the X-Ray transmission through a PCB, resulting in a much lower signal-to-noise ratio and low-quality 3D reconstruction of the PCB sample \cite{asadizanjani2015non}. The combination of the anti-RE techniques for external RE with these techniques for internal RE could provide a holistic solution to protect the system from RE.

\section{Future Research Directions}\label{sec:futureDirections}
Up until this point, we have described the main building blocks of a typical RE framework (see Figure \ref{workflow}(a)), put together to meet the main requirements of an automated approach. 
Moreover, numerous studies introducing and employing various ML and image analysis techniques are reviewed, where the most prominent ones used at various stages of the RE workflow are summarized in Table~\ref{tab:summary}. 
In this respect, the advantages and disadvantages of those algorithms are highlighted and assessed with regard to their applications and performance. 
In fact, the end goal of the entire RE process is to leverage the advantages offered by image analysis and machine learning and eventually perform RE in an automated manner. In addition to obvious benefits, namely the reduction in manpower and costs, automated RE enables a variety of applications, which can further exploit the information provided by the RE. In this context, the following topics and possible future research directions are suggested. 


\begin{enumerate}
\item{Enhanced Hardware Assurance:}\label{sec:enhancedAssurance}
    With automated RE achieved at both of the IC and PCB-levels, it can allow for enhanced levels of assurance, when validating or verifying a design. The detection of defects, design alterations, or IP infringement can be improved to a pixel-level accuracy and performed more efficiently. However, it is not known whether a substantial improvement over the state-of-the-art techniques can be achieved through these advancements. This needs to be explored in future work in this area.   
    
    \vspace{10pt}\item{Qualitative Evaluation of Reverse Engineering: }\label{sec:evaluation} While each block in the RE framework is necessary to achieve a high level of automation, no method has been yet developed in the literature that can evaluate the performance of the RE process at each stage of the framework, let alone the whole process. Defining metrics and criteria for the quality assessment of RE 
    can also provide the ability to explore not only the trade-offs between the quality at various stages, but also their impact on the entire RE process. This can, of course, facilitate possible improvements in efficiency in terms of time, complexity, and required computational resources. 
    
    \vspace{10pt}\item{Benchmarking: } Besides the above point regarding the performance analysis of the algorithms, providing a quantitative comparison between their performance is further challenging. This is a direct consequence of the lack of a comprehensive benchmark dataset that can be used to compare the algorithms. Such a database cannot be built easily due to the complex undertaking involved in preparing instances. This problem can be even more severe for deep learning algorithms, as discussed below.
    
    \vspace{10pt}\item{Reverse Engineering Optimization via Deep Learning:} Due to the lack of a sufficiently large amount of data as well as the variations in data collected across designs, the requirements of effective deep learning models cannot be fulfilled. However, after conducting an RE process, certain stages, e.g., de-noising, via/trace feature extraction, etc., can be optimized via deep learning. In this regard, we can employ synthetic data generation and augmentation, along with using previously extracted RE data as ground truth. 
    This can indeed help to make those stages more time-efficient, when conventional deep neural networks or convolutional neural networks are applied. As this has not been thoroughly explored, the actual extent of such improvement is not known. 
    
        \vspace{10pt}\item{Functional vs. Design Equivalence:}\label{sec:functionalVsParametric}
    Our earlier discussion of image analysis- and machine learning-enabled RE has focused on developing a method to reproduce an IC or, similarly, a PCB as close to the original one as possible. However, there are often scenarios, where only a functionally-equivalent reproduction is necessary rather than a precise design reproduction. As an example, we can refer to an automated generation of a new design built upon some arbitrary standard cells, but with the same functionally as the original one with the foundry-specific standard cells. For PCBs, this can be even more straightforward: producing a PCB, whose connectivity is maintained, but not its specifications, e.g., the trace/via width or distances. Future research should not only focus on such cases, but also go beyond those by considering a broader range of equivalence options, instead of solely functional and design equivalence.  
    
    \vspace{10pt}\item{Missing Data Reconstruction:}\label{sec:missingData}
     Regardless of the modality, one has to deal with the noise in the collected data. The impact of the noise can range from missing regions of the information in a de-processed IC to missing artifacts on a PCB after performing the X-Ray CT. If RE can be automated, it is reasonable to acquire a large amount of data to model particular noise characteristics observed during image acquisition. These models could then be employed to correct noisy data and even reconstruct missing data. While this is well studied in machine learning- and image analysis-related literature, the feasibility and extent to which this can be achieved for IC and PCB RE have yet to be explored. 
    
    \vspace{10pt}\item{Compression Algorithms for RE:}\label{sec:compressionData}
    Another interesting aspect of RE to be taken into account in future work is that the minimum amount of information or data required for a successful reconstruction of images should be collected and stored on a computer system. In addition to speeding up the process of image acquisition, this has the added benefit in terms of space complexity. Unlike common image compression algorithms that attempt to attain the highest possible level of visual quality, new algorithms can be designed or adapted to accurately reconstruct the required features, which leads to a saving of the storage and processing resources. 
\afterpage{
\begin{landscape}
\begin{center}
\scriptsize
\begin{longtable}{|p{0.075\textheight}|
p{0.12\textheight}|
p{0.03\textheight}|
p{0.2\textheight}|
p{0.15\textheight}|
p{0.15\textheight}|
p{0.2\textheight}| p{0.2\textheight}|}
\caption{Summary of the methods applied in the context of RE}
\label{tab:summary}\\
\hline
Article & Main Contribution & Scope & Algorithms & Features Used & Metrics & Evaluation Method & Shortcomings for RE\\
\hline
\endfirsthead
\multicolumn{8}{c}%
{\tablename\ \thetable\ -- \textit{Continued from previous page}} \\
\hline
Article & Main Contribution & Scope & Algorithms & Features Used & Metrics & Evaluation Method & Shortcomings for RE\\
\hline
\endhead
\hline \multicolumn{8}{r}{\textit{Continued on next page}} \\
\endfoot
\hline
\endlastfoot
\cite{lagunovsky1998recognition} & Segmentation & IC & Edge detection & Metal layers & NA & Visual & Optical microscope images\\
\hline
\cite{cheng2018hybrid} & Segmentation & IC & K-means and SVM & Contact and Metal layers & F-score & Ground truth & Only applicable to vias and metal connections \\
\hline
\cite{cheng2018hierarchical, cheng2019global} & Segmentation & IC & K-means, Fuzzy C-Means, SVM & Polysilicon and Contact layers & Intersection over Union (IoU), pixel accuracy & Ground truth & Requires population of shape library \\
\hline
\cite{trindade2018segmentation, masalskis2008reverse, doudkin2005objects} & Segmentation & IC & Spatial and frequency domain filtering & Contact and Metal layers & NA & Visual & Naive applications of image processing\\ 
\hline
\cite{hong2018deep} & Segmentation & IC & DCNN & Contact and Metal layers & False Positive, Negative and semantic errors & Ground truth & Needs 500,000 labelled samples to train\\
\hline
\cite{courbon2019practical} & Extraction of Standard Cell Library & IC & Normalized cross-correlation & Doping layer & NA & Ground truth & Designed for partial RE \\
\hline
\cite{wilson2019novel} & Extraction of Standard Cell Library & IC & Rule based & Contact layer & NA & AES designs with ground truth & Over/Under-segmented cells\\
\hline
\cite{zavadsky2009method, zavadsky2010method} & Localizing standard cells & IC & Template Matching & Conductive layers & NA & NA & Uses approximation to speed up cell localization \\ 
\hline
\cite{kenneth1992automated, ahmed1993integrated} & Netlist generation & IC & Template Matching & All layers & NA & NA & Requires standard cell library\\
\hline
\cite{quijada2018large} & Netlist generation & IC & XGBoost & All layers with pixel intensity, gradient and Hu moments & NA & Ground truth & Needs to be fine tuned for different ICs\\
\hline
\cite{li2012reverse, li2013wordrev, subramanyan2013reverse, baehr2019machine} & High-level description of sub-circuits & IC & Topological analyses, fuzzy structural similarity & Netlist & NA & Ground truth & Uses similarity between known libraries of functional blocks to generate description \\ 
\hline
\cite{schobertgnu, thomasimpact, torrance2009state} & Development of tools for RE & IC & NA & All layers & NA & NA & Assumes availability of some information, e.g., the standard cell library \\
\hline
\cite{vashistha2018trojan, shi2019golden} & IC Trojan detection on active layer & IC & Rule based & Shape of logic cells & NA & Designed on-chip training data & Semi-destructive, need manual polishing \\
\hline
\multicolumn{8}{|r|}{\textit{Continued on next page}} \\
\end{longtable}
\end{center}
\end{landscape}

\pagebreak
\begin{landscape}
\begin{center}
\scriptsize
\begin{longtable}{|p{0.075\textheight}|
p{0.12\textheight}|
p{0.03\textheight}|
p{0.2\textheight}|
p{0.15\textheight}|
p{0.15\textheight}|
p{0.2\textheight}| p{0.2\textheight}|}
\hline
Article & Main Contribution & Scope & Algorithms & Features Used & Metrics & Evaluation Method & Shortcomings for RE\\
\hline
\endfirsthead
\multicolumn{8}{c}%
{\tablename\ \thetable\ -- \textit{Continued from previous page}} \\
\hline
Article & Main Contribution & Scope & Algorithms & Features Used & Metrics & Evaluation Method & Shortcomings for RE\\
\hline
\endhead
\hline \multicolumn{8}{r}{\textit{Continued on next page}} \\
\endfoot
\hline
\endlastfoot
\cite{asadizanjani2017pcb} & Non-Destructive Imaging and Netlist Extraction & PCB & X-Ray CT Imaging and Image Segmentation & X-Ray CT Slices & NA & Dataset with De-Populated Board & Very manual and parameter dependent image processing\\  
\hline
\cite{leta2008computer, chavan2016quality, chaudhary2017automatic, kaur2014detection, tian2014application,dave2016pcb, tang2019online} & PCB Defect Detection and Quality Control & PCB & Image Subtraction\cite{leta2008computer, chavan2016quality, chaudhary2017automatic, kaur2014detection, tian2014application, dave2016pcb} and Deep Learning\cite{tang2019online} & PCB Layer Images & Precision and F-Score & Golden PCB Layer Image & Requires bare board golden layer images\\
\hline
\cite{li2014text} & Text Recognition on PCB Surface & PCB & Optical Character Recognition, Binarization, and Background Estimation & Board surface image & F-Score, Precision, and Recall & Compared against freely available OCR engines OCRAD, Tesseract-OCR, Cuneiform-linux, and GOCR on ground truth & Sub-optimal accuracy \\
\hline
\cite{qiao2018wire} & Trace segmentation & PCB & DCNN and Graph Cuts-based Semantic Segmentation & PCB CT Images & Pixel accuracy, IoU, F1-Score, Precision, Recall & 50 PCB CT test images 
& Dealing with the noise,  
reliance on training data, and variance across designs/imaging.\\ 
\hline
\end{longtable}
\end{center}
\end{landscape}
}

    \vspace{10pt}\item{Cross-Modality Comparison and Evaluation:}\label{sec:crossModality}
    One of the biggest challenges facing us, when evaluating the performance of an RE process, is to compare results across different modalities. Irrespective of the modality (i.e., SEM, Optical, X-Ray CT), the final output of the RE framework should be compared to either a software-provided golden design in a digital format or a PCB schematic. This also requires the exploration of new techniques that can incorporate various factors including different characteristics of the data, the impact of the noise, and variations across the modalities. 
    
    \vspace{10pt}\item{Countermeasures against Advanced RE-based Attacks: }\label{sec:mlcvCounter}
While the techniques mentioned in earlier sections are useful to stop an attacker enjoying the advantages of the current, commonly applied RE framework, this cannot be guaranteed in the future. In particular, to deal with adversaries that can conduct RE, enhanced through the adoption of machine learning and image analysis algorithms, the designer has to predict the risk of such emerging attacks. 
To this end, it is crucial to \emph{estimate} the amount of effort that the attacker has to put and design measures to make the RE process significantly less effective and inefficient. 
Unfortunately, such estimations cannot be carried out in a straightforward manner. 
Therefore, new research directions regarding counter automated-RE development can be of great interest for researchers from government, industry, and academia. 
\end{enumerate}
\section{Conclusion}\label{sec:concl}
In this tutorial, we have comprehensively discussed the challenges associated with RE of ICs and PCBs. It has been observed that for hardware trust and assurance, even though existing, well-known methods (e.g., functional analysis) can be considered useful, the challenges of such methods rise in proportion to the complexity of the IC. The aim of the design in the semiconductor industry is to move towards a higher performance and efficiency and the upward trend of complexity can remain intact for the years to come. Hence, the need for effective RE becomes greater than ever before. 

From the imaging perspective, it has been noted that the primary challenge to be faced by an effective RE-based method is the lack of understanding of the nature of the noise in imaging modalities used for RE, e.g., SEM and CEM. This can be augmented by the techniques used to pre-process the IC for imaging, such as decapsulation and delayering. 
Furthermore, the time spent to acquire high-quality images with a reduced noise level makes RE infeasible for ICs employing today's technology nodes. 

Our overview of machine learning going hand in hand with RE has demonstrated 
the need for quantifying the amount of useful information in each layer. In addition, the effect of counter-RE on machine learning-enabled techniques has been further investigated. Finally, the high variability of features between the layers of different ICs and the lack of high-quality datasets addressing this issue have been observed and regarded as an obstacle to the employment of deep learning for RE.

While relevant literature regarding PCB RE is not rich, there has been a clear interest in taking advantage of image analysis and machine learning algorithms for quality assurance in PCB manufacturing processes. These approaches typically involve image subtraction or (to some extent) building models for detecting defect/anomaly during the manufacturing process. Nevertheless, algorithms implemented for these purposes are not sufficient as they cannot resolve challenges with the external or internal PCB RE.  As a prime example, external PCB RE requires robust, illumination-invariant algorithms to be developed, which ensures effective, high-quality component extraction and analysis. In addition to the intensity inhomogeneity, algorithms designed to deal with internal PCB RE must also be robust to blurring artifacts caused by the X-ray, high-z materials, and aliasing from neighboring layers. Furthermore, due to the wide variety of designs across technologies and the lack of representative datasets, a generalization of the results can pose a serious problem to RE. 

Finally, we believe that this tutorial paves the way for the necessary broad discussion on the above issues and how hardware trust and assurance can benefit greatly from RE.

\begin{acks}
This paper is based upon work supported by Cisco, AFOSR under award No. FA9550-14-1-0351, National Science Foundation under grant No.1821780, and National Science Foundation Graduate Research Fellowship under Grant Nos. 1315138 and 1842473.
\end{acks}

\bibliographystyle{ACM-Reference-Format}
\small
\bibliography{accessbib}

%
%
%

\end{document}